\documentclass[hyper,a4paper]{JHEP}
\usepackage{epsfig}
\usepackage{latexsym}
\usepackage{graphicx}
\usepackage{amsmath}
\usepackage{amsfonts}   
\usepackage{amssymb}    

\def\ee{e^+e^-\to h^0Z}

\def\lsim{\raise0.3ex\hbox{$\;<$\kern-0.75em\raise-1.1ex\hbox{$\sim\;$}}}\def\gsim{\raise0.3ex\hbox{$\;>$\kern-0.75em\raise-1.1ex\hbox{$\sim\;$}}}

\newcommand{\captions}{\sf\caption}
\def    \be            {\begin{equation}}
\def    \ee            {\end{equation}}
\def    \bea           {\begin{eqnarray}}
\def    \eea           {\end{eqnarray}}

\def\sw2{sin^2 \theta_w}

\def\diag{\mathrm{diag}}
\def\I{\mathrm{i}}

\def\a^tau{\alpha_{\tau}}

\def\beq{\begin{equation}}
\def\eeq{\end{equation}}
\def\beqa{\begin{eqnarray}}
\def\eeqa{\end{eqnarray}}

\title{
Neutrino Physics and Spontaneous CP Violation in the $\mu\nu$SSM
 }
\author{Javier Fidalgo\\
        Departamento de F\'{\i}sica Te\'{o}rica
        and Instituto de F\'{\i}sica Te\'{o}rica UAM/CSIC,\\
        Universidad Aut\'{o}noma de Madrid, Cantoblanco,
        28049 Madrid, Spain\\
        E-mail: \email{fidalgo@delta.ft.uam.es}}
\author{Daniel E. L\'{o}pez-Fogliani\\
       Department of Physics and Astronomy, University of Sheffield,\\
        Sheffield S3 7RH, England\\
        E-mail: \email{d.lopez@sheffield.ac.uk}}
\author{Carlos Mu\~noz\\
        Departamento de F\'{\i}sica Te\'{o}rica
        and Instituto de F\'{\i}sica Te\'{o}rica UAM/CSIC,\\
        Universidad Aut\'{o}noma de Madrid, Cantoblanco,
        28049 Madrid, Spain\\
        E-mail: \email{carlos.munnoz@uam.es}}
\author{Roberto Ruiz de Austri\\
        Instituto de F\'{\i}sica Corpuscular IFIC-UV/CSIC\\
        Universidad de Valencia,Valencia, Spain \\
        E-mail: \email{rruiz@ific.uv.es}}
\abstract{\small 
The $\mu\nu$SSM provides a solution to the
$\mu$ problem of the MSSM and explains the origin of neutrino masses by simply
using right-handed neutrino superfields.
We have completed the analysis of the vacua in this model,
studying the possibility of spontaneous CP violation 
through complex Higgs and sneutrino vacuum expectation values.
As a consequence of this process, a complex MNS matrix can be present.
Besides, we have discussed the neutrino physics 
and the associated electroweak seesaw mechanism in the $\mu\nu$SSM,
including also phases. Current data on neutrino masses and mixing angles can easily be reproduced.
}
\preprint{
\rightline{FTUAM 09/06, IFT-UAM/CSIC-09-20, April 2009}
}
\begin{document}
\section{Introduction}
Although the minimal supersymmetric extension of the standard 
model (MSSM) reveals as a solution to the hierarchy problem, 
we still remain puzzled about the origin of the $\mu$-term in 
the superpotential, known as the $\mu$-problem \cite{muproblem}. 
On the other hand, the fact that neutrinos are not massless \cite{experiments} 
suggests that the MSSM is incomplete. 
Motivated by these two facts, the
''$\mu$ from $\nu$'' supersymmetric standard model
($\mu\nu$SSM) \cite{MuNuSSM,MuNuSSM2,MuNuSSM0}, which relies on the
existence of right-handed neutrinos, 
arises as an alternative to the MSSM, 
providing a solution to the
$\mu$-problem 
and explaining the origin of neutrino masses.

In particular, the superpotential of the $\mu\nu$SSM contains, in addition to the
usual Yukawa couplings for quarks and charged leptons,
Yukawa couplings for neutrinos
$\hat H_u\,  \hat L \, \hat \nu^c$, terms of the type
$\hat \nu^c \hat H_d\hat H_u$ 
producing an effective  $\mu$ term through right-handed sneutrino
vacuum expectation values (VEVs),
and also terms of the type $\hat \nu^c \hat \nu^c \hat \nu^c$  
avoiding the existence of a Goldstone boson and
contributing to generate
effective Majorana masses for neutrinos at the electroweak scale.
Actually, the explicit breaking of R-parity in this model 
by the above terms produces the mixing of neutralinos with
left- and right-handed neutrinos, and as a consequence a generalized matrix of the
seesaw type that gives rise at tree level to three
light eigenvalues corresponding to neutrino 
masses \cite{MuNuSSM}.

Following this proposal, several papers have studied different aspects of the $\mu\nu$SSM.
In \cite{MuNuSSM2}
the parameter space of the $\mu\nu$SSM was analyzed
in detail, studying the viable regions which avoid false minima and tachyons, as well as fulfill the Landau pole constraint. The structure of the mass matrices, and the
associated particle spectrum was also computed, paying
special attention to the mass of the lightest Higgs.
In \cite{Ghosh:2008yh} neutrino masses and mixing angles were discussed, as well as the decays of the lightest neutralino
to two body ($W$-lepton) final states.
The correlations of the decay branching ratios with the neutrino mixing angles were studied as another possible test of the $\mu\nu$SSM at the LHC.
The phenomenology of the $\mu\nu$SSM was also studied
in \cite{Hirsch}, particularized for one and two generations
of right-handed sneutrinos, and taking into account all 
possible final states when studying the decays of the lightest neutralino. 
Possible signatures that might allow to distinguish this model from other R-parity breaking models were discussed qualitatively in the last two papers. 
Let us finally mention that
terms of the type
$\hat \nu^c \hat H_d \hat H_u$ and $\hat \nu^c \hat \nu^c \hat \nu^c$
were also analysed as sources of the observed baryon asymmetry
in the Universe \cite{vallle} and of neutrino masses and bilarge mixing \cite{sri}, respectively.


The goal of this work is twofold; first, we complete the analysis of the vacua of the $\mu\nu$SSM 
presented in \cite{MuNuSSM2}, studying spontaneous CP violation (SCPV) of the tree-level neutral scalar potential. 
In particular, we explore CP violation in the lepton sector
and show how phases for the tree-level Maki-Nakagawa-Sakata 
matrix (MNS) \cite{mns} may arise due to the fact that the minimum of the scalar potential with real parameters has complex VEV solutions.
Second, we discuss neutrino physics and the seesaw mechanism in the $\mu\nu$SSM, including also phases.

Let us recall that, although there is evidence for CP violation in the quark 
sector of the standard model, there are not experimental traces of it in the leptonic part. 
CP can be explicitly broken through 
complex parameters in the Lagrangian or can arise spontaneously in a CP conserving Lagrangian 
(e.g. with all the parameters being real) through complex VEVs.
Although the standard model as well as the MSSM do not allow 
for SCPV, in more complicated models both sources of CP violation, complex parameters and complex VEVs, 
could be present.

Concerning the quark sector, a recent study argues that the 
Cabibbo-Kobayashi-Maskawa (CKM) matrix 
is likely 
complex \cite{botella}. This conclusion is supported 
by the measurement of the unitarity triangle angle $\gamma$ 
by BaBar and Belle collaborations \cite{babar, belle}. 
This evidence of a complex CKM matrix has ruled out Next-to-MSSM (NMSSM)-like models with 
SCPV (see e.g. \cite{NMSSM with SCPV rulled-out by Botella}) for being the 
entire source of CP violation in the quark sector, since the CKM matrix in 
such models is real. Thus complex parameters are necessary
in the quark sector.
Given the structure of the $\mu\nu$SSM, this fact also holds for this model.
On the other hand, as mentioned above, we will show that SCPV can be generated
in the leptonic sector of the $\mu\nu$SSM, as well as phases for the MNS matrix.

One argument in favor of the presence of SCPV
at the Lagrangian level is that, if the determinant of the quark mass 
matrix is real, it leads to a solution to the strong CP problem 
\cite{theta}. 
Extensions of the MSSM having this property, have been extensively studied 
in the literature (see e.g. \cite{costa}). In those scenarios, the quark 
sector of the model is extended in such a way that the effective 
$3 \times 3$ CKM matrix is complex whereas the determinant of the quark matrix is real. 

Other authors have extended the Higgs sector of the models, leading to SCPV with a
complex CKM matrix \cite{higgsraros}. 
Last but not least, in supersymmetric (SUSY) models with both CP and Peccei-Quinn 
symmetries, SCPV can be used as a solution to the SUSY phase problem \cite{phases}.


Regarding extensions of the $\mu\nu$SSM, the SCPV 
scenario with a complex CKM matrix can be accomplished by adding two more families of Higgs doublets.
In this case the model would contain three families of matter and Higgs fields. This possibility is well motivated phenomenologically, since the potential problem of flavor changing neutral currents
can be avoided \cite{EscMunTei}.
In addition, having three Higgs families is favored 
in some string scenarios \cite{EscMunTei1}.
Indeed, extensions of the quark sector of the model can also be studied,
without altering the results here presented.


What we want to point out in this work is that SCPV 
is possible in the simplest version of the $\mu\nu$SSM, 
i.e. with only one family of Higgs doublets, and therefore it is worth 
studying its consequences. 
Following this philosophy, 
the paper is organized as follows. Section \ref{section:vacuua} is devoted 
to complete the analysis of the vacuum of the
$\mu\nu$SSM  started in \cite{MuNuSSM2}, 
including SCPV solutions. In Section \ref{section:neutrino} 
we examine the seesaw mechanism as the origin of neutrino masses 
and mixing angles in the model. In Section \ref{section:results} we carry out 
a detailed numerical analysis of the tree-level neutral scalar 
potential, showing explicitly that SCPV solutions are possible, and discussing their implications on the 
neutrino sector of the model. Finally, the conclusions are left for 
Section \ref{section:conclusion}.
Minimization equations of the model and an approximate
analytical formula for neutrino masses are given in the Appendices.

\section{Complex VEVs in the $\mu\nu$SSM \label{section:vacuua}}

The superpotential of the $\mu \nu$SSM introduced in \cite{MuNuSSM} is 
given by
\begin{align}\label{superpotential}
W = &
\ \sum_{a,b}  \sum_{i,j} \left[ \epsilon_{ab} \left(
Y_{u_{ij}} \, \hat H_u^b\, \hat Q^a_i \, \hat u_j^c +
Y_{d_{ij}} \, \hat H_d^a\, \hat Q^b_i \, \hat d_j^c +
Y_{e_{ij}} \, \hat H_d^a\, \hat L^b_i \, \hat e_j^c +
Y_{\nu_{ij}} \, \hat H_u^b\, \hat L^a_i \, \hat \nu^c_j 
\right) \right]
\nonumber\\
& 
-\sum_{a,b} \sum_i \epsilon{_{ab}} \lambda_{i} \, \hat \nu^c_i\,\hat H_d^a \hat H_u^b
+
\sum_{i,j,k}\frac{1}{3}
\kappa{_{ijk}} 
\hat \nu^c_i\hat \nu^c_j\hat \nu^c_k\,  ,
\end{align}
where we take $\hat H_d^T=(\hat H_d^0, \hat H_d^-)$, 
$\hat H_u^T=(\hat H_u^+, \hat H_u^0)$, $\hat Q_i^T=(\hat u_i, \hat d_i)$, 
$\hat L_i^T=(\hat \nu_i, \hat e_{L_i})$, 
$i,j,k=1,2,3$ are family indices, the 3$\times$3 matrices $Y$ are 
dimensionless Yukawa couplings, $a,b=1,2$ are $SU(2)_L$ indices and 
$\epsilon_{12}=1$. As mentioned in the Introduction, in addition to the MSSM Yukawa couplings 
for quarks and 
charged leptons, the $\mu\nu$SSM superpotential contains Yukawa couplings for 
neutrinos, and  two additional type of terms involving the Higgs 
doublet superfields, $\hat H_d$ and $\hat H_u$ and the three right-handed neutrino 
superfields, $\hat \nu^c_i$, with the dimensionless vector coupling 
$\lambda$ and the totally symmetric tensor $\kappa$.

As discussed in \cite{MuNuSSM}, when the scalar components of the 
superfields $\hat\nu^c_i$, denoted by $\tilde\nu^c_i$, acquire VEVs 
of the order of the electroweak scale, an effective interaction 
$\mu \hat H_1 \hat H_2$ is generated through the fifth term in Eq.
(\ref{superpotential}), with  
$\mu\equiv \lambda_i \langle \tilde \nu^c_i \rangle$. 
The last type of terms in Eq. (\ref{superpotential}) is allowed by 
all symmetries, and avoids the presence of an unacceptable Goldstone 
boson associated to a global $U(1)$ symmetry.
In addition, it generates effective Majorana masses for neutrinos 
at the electroweak scale. These two type of terms break explicitly 
$R$-parity and lepton number.

Working in the framework of gravity mediated supersymmetry breaking, 
the Lagrangian  $\mathcal{L}_{\text{soft}}$ 
is given by:
\begin{eqnarray}
-\mathcal{L}_{\text{soft}} & =&
 \sum_{i,j} \left[\sum_a m_{\tilde{Q}_{ij} }^2\, \tilde{Q^a_i}^* \, \tilde{Q^a_j}
+m_{\tilde{u}_{ij}^c}^{2} 
\, \tilde{u^c_i}^* \, \tilde u^c_j
+m_{\tilde{d}_{ij}^c}^2 \, \tilde{d^c_i}^* \, \tilde d^c_j
+\sum_a m_{\tilde{L}_{ij} }^2 \, \tilde{L^a_i}^* \, \tilde{L^a_j} \right. \nonumber \\
&+ &
\left. m_{\tilde{e}_{ij} ^c}^2 \, \tilde{e^c_i}^* \, \tilde e^c_j +  m_{\tilde{\nu}_{ij}^c}^2 \,\tilde{{\nu}^c_i}^* \tilde\nu^c_j  \right]
\nonumber \\
&+ &
\sum_a \left[ m_{H_d}^2 \,{H^a_d}^*\,H^a_d + m_{H_u}^2 \,{H^a_u}^* H^a_u \right]
\nonumber \\
&+&
\sum_{a,b} \sum_{i,j}\epsilon_{ab} \left[
(A_uY_u)_{ij} \, H_u^b\, \tilde Q^a_i \, \tilde u_j^c +
(A_dY_d)_{ij} \, H_d^a\, \tilde Q^b_i \, \tilde d_j^c +
(A_eY_e)_{ij} \, H_d^a\, \tilde L^b_i \, \tilde e_j^c 
\right.
\nonumber \\
&+&
\left.
(A_{\nu}Y_{\nu})_{ij} \, H_u^b\, \tilde L^a_i \, \tilde \nu^c_j 
+ \text{c.c.}
\right] 
\nonumber \\
&+&
\left[-\sum_{a,b}\sum_i \epsilon_{ab} (A_{\lambda}\lambda)_{i} \, \tilde \nu^c_i\, H_d^a  H_u^b
+
\sum_{ijk}\frac{1}{3}
(A_{\kappa}\kappa)_{ijk} \, 
\tilde \nu^c_i \tilde \nu^c_j \tilde \nu^c_k\
+ \text{c.c.} \right]
\nonumber \\
&-&  \frac{1}{2}\, \left(M_3\, \tilde\lambda_3\, \tilde\lambda_3+M_2\,
  \tilde\lambda_2\, \tilde
\lambda_2
+M_1\, \tilde\lambda_1 \, \tilde\lambda_1 + \text{c.c.} \right) .
\label{2:Vsoft}
\end{eqnarray}

In addition to terms from $\mathcal{L}_{\text{soft}}$, the 
tree-level scalar potential receives the $D$ and $F$ term
contributions also computed in \cite{MuNuSSM}.
In the following we will suppose that CP is a good symmetry of the model, 
taking all the parameters in the neutral scalar potential real and assuming 
that CP is only violated by the VEVs of the scalar 
fields
%
\begin{equation}\label{2:vevs}
\langle H_d^0 \rangle = e^{i\varphi_{v_d}} \: v_d \, , \quad
\langle H_u^0 \rangle = e^{i \varphi_{v_u}}\:v_u \, , \quad
\langle \tilde \nu_i \rangle = e^{\varphi_{\nu_i}}\:\nu_i \, , \quad
\langle \tilde \nu_i^c \rangle = e^{\varphi_{\nu^c_i}}\:\nu_i^c \,.
\end{equation}
We then obtain for the tree-level neutral scalar potential, 
\begin{equation}
V^0 = V_{\text{soft}} + V_D  +  V_F\ , 
\label{finalpotential}
\end{equation}
where
\begin{eqnarray}
V_{\text{soft}} &=& m_{H_d}^{2}v_{d}v_{d}+m_{H_u}^{2}v_{u}v_{u}+
\sum_{i,j} m_{\tilde{L}_{ij} }^2 \, {\nu_i} \, {\nu_j} \cos(\chi_i-\chi_j) +
\sum_{i,j} m_{\tilde{\nu}^c_{ij}}^{2}{\nu}^c_{i}{\nu}^c_{j} \cos(\varphi_{\nu^c_i}
-\varphi_{\nu^c_j})
\nonumber\\
&-&2\sum_{i} (A_{\lambda} \lambda)_{i}{\nu}^c_{i}v_{d}v_{u} \cos(\varphi_{v} 
+\varphi_{\nu^c_i})
+\sum_{i,j,k} \frac{2}{3} 
 { ( A_{\kappa} \kappa)_{ijk}{\nu}^c_{i}{\nu}^c_{j}{\nu}^c_{k} 
  cos(\varphi_{\nu^c_i}+\varphi_{\nu^c_j}+\varphi_{\nu^c_k})}
\nonumber \\ 
&+& 2\sum_{i,j} (A_{\nu} Y_{\nu})_{ij} v_{u}{\nu}_{i}{\nu}^{c}_{j} 
\cos(\chi_i+\varphi_{\nu^c_j})\ ,
\label{Vsoft}
\end{eqnarray}
%
\begin{eqnarray}
V_D=\frac{G^2}{8}\left(\sum_i {\nu}_{i}{\nu}_{i} 
                       + v_{d}v_{d}-v_{u}v_{u}\right)^{2}\ , 
\label{V_D}
\end{eqnarray}
with $G^2\equiv g_{1}^{2}+g_{2}^{2}$, and
\begin{eqnarray}
V_{F} &=&\sum_{i}(\lambda_{i})^2v^2_{d}v^2_{u} + 
\nonumber\\
&+&\sum_{i,j}\lambda_{i}\lambda_{j}v_{d}^2{\nu}^c_{i}{\nu}^{c}_{j} 
  \cos(\varphi_{\nu^c_i}-\varphi_{\nu^c_j})
+\sum_{i,j}\lambda_{i}\lambda_{j}v_{u}^2{\nu}^c_{i}{\nu}^{c}_{j}
 \cos(\varphi_{\nu^c_i}-\varphi_{\nu^c_j})
\nonumber\\
&+&\sum_{i,j,k,l}\sum_{m}
\kappa_{imk}\kappa_{lmj}{\nu}^c_{i}{\nu}^{c}_{j}{\nu}^c_{k}{\nu}^{c}_{l}
 \cos(\varphi_{\nu^c_i}+\varphi_{\nu^c_j}-\varphi_{\nu^c_k}-\varphi_{\nu^c_l})
\nonumber\\
&+&2 \, 
\left[-\sum_{i,j}\sum_{k}
\kappa_{ikj}\lambda_{k}v_{d}v_{u}{\nu}^c_{i}{\nu}^c_{j}\cos(\varphi_{\nu^c_i}+
 \varphi_{\nu^c_j}-\varphi_v)\ 
\right. \nonumber\\
&+&\sum_{i,j,k}\sum_{l}
Y_{\nu_{jl}}\kappa_{ilk}v_{u}{\nu}_{j}{\nu}^c_{i}{\nu}^c_{k}\cos(\varphi_{\nu^c_i}
+\varphi_{\nu^c_k}-\chi_j)
\nonumber \\
&-&\sum_{i,j,k}Y_{\nu_{ij}}\lambda_{k}v_{d}{\nu}_{i}{\nu}^c_{j}{\nu}^c_{k} 
 \cos(\chi_i+\varphi_{\nu^c_j}-\varphi_{\nu^c_k}-\varphi_{v})\ 
\nonumber\\
&-&
\left. \sum_{i}\sum_{j}Y_{\nu_{ij}}\lambda_{j}v_{d}v^2_{u}{\nu}_{i}
\cos(\varphi_{v}-\chi_i)
\right]
\nonumber\\
&+&\sum_{i,j,k,l}Y_{\nu_{ij}}Y_{\nu_{kl}}{\nu}_{i}{\nu}^c_{j}{\nu}_{k}{\nu^{c}_{l}} 
\cos(\chi_i-\chi_k+\varphi_{\nu^c_j}-\varphi_{\nu^c_l})
\nonumber \\
&+&\sum_{i,j}\sum_{k}Y_{\nu_{ik}}Y_{\nu_{jk}}v^2_{u}{\nu}_{i}{\nu}_{j}
\cos(\chi_i-\chi_j)
\nonumber\\
&+&\sum_{i,j}\sum_{k}Y_{\nu_{ki}} Y_{\nu_{kj}}v^2_{u} {\nu}^c_{i}{\nu^{c}_{j}}
\cos(\varphi_{\nu_i^c}-\varphi_{\nu_j^c})\ . \label{V_F}
\end{eqnarray}
We observe that in the potential there are seven independent phases, 
and we have defined them as
\begin{equation}
\varphi_v=\varphi_{v_u}+\varphi_{v_d}\; , \; \;\;
\chi_i=\varphi_{\nu_i}+\varphi_{v_u}\; , \;\;\; \varphi_{\nu^c_i}. 
\end{equation}

Now one can derive the fifteen minimization conditions 
with respect to the moduli $v_d$, $v_u$, $\nu_i^c$, $\nu_i$, and phases
$\varphi_v$, 
$\chi_i$, and $\varphi_{\nu^c_i}$\ . 
These are written in Appendix A.
Finding minima requires the solutions of equations (\ref{8vd}--\ref{7n}). 
A standard way to obtain this is 
to give the values of the cosines of the phases in terms of the moduli, using the triangle 
method \cite{branco,masip,masip2} for the equation of the 
phases,
and then substitute the expressions in the
minimum equations for the moduli, 
solving them numerically. 
This method permits to demonstrate the existence at tree level of only real minima 
in several models. This is for example the case of the NMSSM \cite{romao86}, and the
MSSM with extra doublets. The latter result has been proved for 
the MSSM with an extra pair of Higgs doublets \cite{masip} (the so called 
4D model), the bilinear R-parity violation model (analogous to a 5D model because of the VEVs of the left-handed sneutrinos), and the MSSM with two extra pair of Higgs doublets (6D model) \cite{masip2}.

Another way of finding minima consists of using as input the phases and solve the 
fifteen equations to fix the variables that are linear in these equations, 
as it is the case of some of the soft terms. This is the procedure that we will follow
in Section~4.

A simple way to prove the existence of CP violating minima in the 
$\mu\nu$SSM is using the results of Ref. \cite{masip2}, where the authors prove 
that SUSY scenarios for SCPV require singlets. In particular, they found that, if the singlets do not introduce dimensional parameters in the superpotential (i.e. no linear or bilinear terms), the MSSM extended with two gauge singlets would be the minimal SUSY model where
CP violation can be generated spontaneously. 
Since that model is a limiting
case
of the $\mu\nu$SSM with vanishing neutrino Yukawa couplings $Y_{\nu_{ij}}=0$, 
$\lambda_3=0$,
and 
$\kappa_{333}=\kappa_{322}=\kappa_{332}=\kappa_{311}=\kappa_{331}=\kappa_{123}=0$,
this would prove that the $\mu\nu$SSM can break CP spontaneously.
Let us remark that, 
since in the  $\mu\nu$SSM one is using a seesaw at the electroweak scale, the $Y_{\nu_{ij}}$ have to be very small compared with 
the other parameters \cite{MuNuSSM}, and as a consequence 
the neutral scalar 
potential can be understood as a deformation of the MSSM extended with three
gauge singlets.
Although 
there is no literature about general solutions that break CP spontaneously 
in the latter, it is obvious that this model contains the MSSM extended with two singlets as a limiting case 
when  
$\kappa_{333}=\kappa_{322}=
\kappa_{332}=\kappa_{311}=\kappa_{331}=\kappa_{123}=0$,
and $\lambda_3=0$. 
As already mentioned,
SCPV solutions are well known in this case 
\cite{masip2,coreanos}. Thus one could argue that a subset of solutions with neutrino masses different from zero could be obtained deforming the scalar potential of the MSSM extended with three singlets\footnote{Since only mass differences for neutrino masses have been measured, in principle 
two right-handed neutrino supermultiplets are enough to give two 
tree-level masses and also break CP spontaneously. Thus a version of the $\mu\nu$SSM with only two right-handed neutrinos instead of three could be 
formulated. Nevertheless, we will follow the philosophy that the existence of three generations
of all kind of leptons is more natural. 
} through non-zero $Y_{\nu_{ij}}$.


In Sect. \ref{section:results} we will do a thorough numerical analysis showing 
explicitly how SCPV is realizable in the leptonic sector.
Nevertheless, it is worth pointing out here that to find complex solutions
is a non-trivial task compared to the search of real ones.
As we will show, the key of SCPV is on the $(A_\kappa \kappa)_{ijk}$ terms used as inputs. In order to fulfill the minimization equations, the basic 
requirement is that entries different from $(A_\kappa \kappa)_{iii}$ must be allowed.
In addition, these parameters have to be chosen carefully to 
obtain SCPV as a global minimum. 

In the next section we will study the seesaw mechanism in the model 
as the origin of neutrino masses and mixing angles. 

\section{Neutrino masses and mixing angles \label{section:neutrino}}

In the $\mu \nu$SSM the MSSM neutralinos mix with the left- and 
right-handed neutrinos as a consequence of R-parity violation. 
Therefore 
the right-handed neutrinos behave as singlino components of the neutralinos. 
In the basis 
${\chi^{0}}^T=(\tilde{B^{0}},\tilde{W^{0}},\tilde{H_{d}},\tilde{H_{u}}, 
        \nu_{R_i},\nu_{L_i})$ 
the neutralino-neutrino mass matrix was given in
\cite{MuNuSSM, MuNuSSM2} for real VEVs. Considering now
the possibility of complex VEVs the result is given by
\begin{align}
{\cal M}_n=\left(\begin{array}{cc}
M & m\\
m^{T} & 0_{3\times3}
\label{Mneutralinos10x10}
\end{array}\right),
\end{align} 
where the neutralino mass matrix is
{\small \begin{align}
\hspace*{-2.5cm} \hspace{.2mm} M=\hspace{-.2mm}
\left(
\begin{array}{ccccccc}
M_{1} & 0 & -A \langle H_d^0 \rangle^* & A \langle H_u^0 \rangle^* & 0 & 0 & 0\\
0 & M_{2} &B \langle H_d^0 \rangle^* & -B \langle H_u^0 \rangle^* & 0 & 0 & 0\\
-A \langle H_d^0 \rangle^* & B \langle H_d^0 \rangle^* & 0 & 
-\lambda_{i}\langle \tilde \nu_i^c \rangle  & 
-\lambda_{1}\langle H_u^0 \rangle & -\lambda_{2}\langle H_u^0 \rangle & 
-\lambda_{3}\langle H_u^0 \rangle\\
A \langle H_u^0 \rangle^* & -B \langle H_u^0 \rangle^* & 
-\lambda_{i}\langle \tilde \nu_i^c \rangle  & 0 & 
-\lambda_{1}\langle H_d^0 \rangle + Y_{\nu_{i1}}\langle \tilde \nu_i \rangle  & 
-\lambda_{2}\langle H_d^0 \rangle + Y_{\nu_{i2}}\langle \tilde \nu_i \rangle  & 
-\lambda_{3}\langle H_d^0 \rangle + Y_{\nu_{i3}}\langle \tilde \nu_i \rangle \\
0 & 0 & -\lambda_{1}\langle H_u^0 \rangle & 
-\lambda_{1}\langle H_d^0 \rangle + Y_{\nu_{i1}}\langle \tilde \nu_i \rangle  & 
2\kappa_{11j}\langle \tilde \nu_j^c \rangle  & 
2\kappa_{12j}\langle \tilde \nu_j^c \rangle  & 
2\kappa_{13j}\langle \tilde \nu_j^c \rangle \\
0 & 0 & -\lambda_{2}\langle H_u^0 \rangle & 
-\lambda_{2}\langle H_d^0 \rangle + Y_{\nu_{i2}}\langle \tilde \nu_i \rangle  & 
2\kappa_{21j}\langle \tilde \nu_j^c \rangle  & 
2\kappa_{22j}\langle \tilde \nu_j^c \rangle  & 
2\kappa_{23j}\langle \tilde \nu_j^c \rangle \\
0 & 0 & -\lambda_{3}\langle H_u^0 \rangle & 
-\lambda_{3}\langle H_d^0 \rangle + Y_{\nu_{i3}}\langle \tilde \nu_i \rangle  & 
2\kappa_{31j}\langle \tilde \nu_j^c \rangle  & 
2\kappa_{32j}\langle \tilde \nu_j^c \rangle  & 
2\kappa_{33j}\langle \tilde \nu_j^c \rangle \end{array}
\right), \label{Mneutralinos7x7}
\end{align}
}
with $A = \frac{G}{\sqrt{2}} \sin\theta_W$, 
$B = \frac{G}{\sqrt{2}} \cos\theta_W$, and
\begin{align}
m^{T}=\left(\begin{array}{ccccccc}
-\frac{g_{1}}{\sqrt{2}}\langle \tilde \nu_1 \rangle^*  & 
\frac{g_{2}}{\sqrt{2}}\langle \tilde \nu_1 \rangle^*  & 
0 & 
Y_{\nu_{1i}}\langle \tilde \nu_i^c \rangle  & 
Y_{\nu_{11}}\langle H_u^0 \rangle & 
Y_{\nu_{12}}\langle H_u^0 \rangle & Y_{\nu_{13}}\langle H_u^0 \rangle\\
-\frac{g_{1}}{\sqrt{2}}\langle \tilde \nu_2 \rangle^*  & 
\frac{g_{2}}{\sqrt{2}}\langle \tilde \nu_2 \rangle^*  & 
0 & 
Y_{\nu_{2i}}\langle \tilde \nu_i^c \rangle  & 
Y_{\nu_{21}}\langle H_u^0 \rangle & 
Y_{\nu_{22}}\langle H_u^0 \rangle & 
Y_{\nu_{23}}\langle H_u^0 \rangle\\
-\frac{g_{1}}{\sqrt{2}}\langle \tilde \nu_3 \rangle^*  & 
\frac{g_{2}}{\sqrt{2}}\langle \tilde \nu_3 \rangle^*  & 
0 & 
Y_{\nu_{3i}}\langle \tilde \nu_i^c \rangle  & 
Y_{\nu_{31}}\langle H_u^0 \rangle & 
Y_{\nu_{32}}\langle H_u^0 \rangle & 
Y_{\nu_{33}}\langle H_u^0 \rangle\end{array}\right). \label{m7x3} \end{align}
For simplicity the summation convention on repeated indices was used in the above two equations.
The matrix (\ref{Mneutralinos10x10}) is of the seesaw type giving rise 
to the neutrino masses which have to be very small. 
This is the case since the entries of the matrix $M$ are much
larger than the ones in the matrix $m$.
Notice in this respect that the entries of $M$ are of the order of the 
electroweak scale while the ones in $m$ are of the order 
of the Dirac masses for the neutrinos \cite{MuNuSSM, MuNuSSM2}. Therefore 
in a first approximation the effective neutrino mixing mass matrix 
can be written as
\begin{equation}
m_{eff} = -m^T \cdot M^{-1} \cdot m. \label{eff}
\end{equation}
Because $m_{eff}$ is symmetric and $m^{\dag}_{eff} m_{eff} $ is Hermitian, 
one can diagonalize them by a unitary transformation
\begin{equation}
U_{MNS}^T m_{eff} U_{MNS} = diag(m_{\nu_1}, m_{\nu_2}, m_{\nu_3}),
\label{nudagnudiag}
\end{equation}
\begin{equation}
U_{MNS}^\dag m_{eff}^{\dag} m_{eff}  U_{MNS} = diag(m_{\nu_1}^2, m_{\nu_2}^2, 
m_{\nu_3}^2).
\label{nudiag}
\end{equation}

In Appendix B, Eq. (\ref{Analytical approximate effective neutrino mass matrix}), we 
present an approximate analytical expression for the effective neutrino mass 
matrix of the $\mu \nu$SSM with SCPV, neglecting all the terms containing $Y_{\nu}^2 \nu^2$, $Y_{\nu}^3 \nu$ and $Y_{\nu} \nu^3$ in Eq. (\ref{eff}) due to the smallness 
of $Y_{\nu}$ and $\nu$ \cite{MuNuSSM}. 
In the limit of vanishing phases $\varphi_{v_u}=
\varphi_{v_d}=\varphi_{\nu^c_i}=\varphi_{\nu_i}=0 $,
Eq. (\ref{Analytical approximate effective neutrino mass matrix}) is reduced 
to Eq. (\ref{Formula analitica real}). This is the formula
that we will use in the following, in order 
to have a qualitative idea of how the seesaw mechanism works in this model.

Let us first rewrite the expression (\ref{Formula analitica real}) in the 
following form:
\begin{eqnarray}
\label{Formula analitica real reescrita}
& (m_{eff|real})_{ij} \simeq \frac{v_u^2}{6\kappa \nu^c}Y_{\nu_i}Y_{\nu_j}
                     \left(1-3 \, \delta_{ij}\right)  
&  -\frac{1}{2M_{eff}}\left[\nu_i \nu_j+\frac{v_d\left(Y_{\nu_i}\nu_j
   +Y_{\nu_j}\nu_i\right)}{3\lambda}
   +\frac{Y_{\nu_i}Y_{\nu_j}v_d^2}{9\lambda^2 }\right]\ ,
   \nonumber\\
  \end{eqnarray}     
with
\begin{eqnarray}
 M_{eff}\equiv M \left[1-\frac{v^2}{2M \left(\kappa \nu^{c^2}+\lambda v_u v_d\right)
        \ 3 \lambda \nu^c}\left(2 \kappa \nu^{c^2} \frac{v_u v_d}{v^2}
        +\frac{\lambda v^2}{2}\right) \right]\ ,
\end{eqnarray}
which coincides with the result in \cite{Ghosh:2008yh}, where the possibility 
of obtaining an adequate seesaw with diagonal Yukawa couplings was also pointed out.
Here
$v^2=v_u^2+v_d^2+\sum_i\nu_i^2 \approx v_u^2+v_d^2$ 
with $v\approx 174$ GeV
has been used, since $\nu_i <<v_u,v_d$ \cite{MuNuSSM},
and let us recall that we are also using couplings
$\lambda_i \equiv \lambda$, 
a tensor $\kappa$ with terms $\kappa_{iii} \equiv \kappa_i \equiv \kappa$ and vanishing otherwise,
diagonal Yukawa couplings $Y_{\nu_{ii}} \equiv Y_{\nu_i}$, 
VEVs $\nu_i^c\equiv\nu^c$,
and $\frac{1}{M} = \frac{g_1^2}{M_1}+\frac{g_2^2}{M_2}$.

In the limit where gauginos are very heavy and decouple
(i.e. $M \rightarrow \infty$), Eq. (\ref{Formula analitica real reescrita}) 
reduces to
\begin{eqnarray}
(m_{eff|real})_{ij} \simeq 
         \frac{v_u^2}{6 \, \kappa \nu^c}Y_{\nu_i}Y_{\nu_j}
\left(1-3 \, \delta_{ij}\right).
\label{Not Ordinary see-saw}
\end{eqnarray}
It is interesting to note that in contrast with the ordinary seesaw 
(i.e. generated only through the mixing between left- and right-handed 
neutrinos), where the case of diagonal Yukawas would give rise to a diagonal mass matrix of the form 
\begin{eqnarray}
(m_{eff|ordinary\ seesaw})_{ij} \simeq 
              \frac{-v_u^2 Y_{\nu_i}Y_{\nu_j}\delta_{ij}}{2 \, \kappa \nu^c},
\end{eqnarray} 
in this case we have an extra contribution given by the 
first term of Eq. (\ref{Not Ordinary see-saw}). 
This is due to the effective mixing of the right-handed neutrinos and Higgsinos in this limit, and produces off-diagonal entries in the mass matrix. Besides, when right-handed neutrinos 
are also decoupled (i.e. $\nu^c \to \infty $), the neutrino masses 
are zero as corresponds to the case of a seesaw with only Higgsinos.

Another observation is that,
independently on the nature of the lightest neutralino, Higgsino-like or $\nu^c$-like or even a mixture of them (recall that the $\nu^c$ can be interpreted also as the singlino component of the neutralino 
since R-parity is broken), the form of the effective neutrino mass matrix 
is the same when the gauginos are decoupled, as given by
(\ref{Not Ordinary see-saw}).

Another limit which is worth discussing is 
$\nu^c \rightarrow \infty$. Then, Eq. 
(\ref{Formula analitica real reescrita}) reduces to the form 
\bea
(m_{eff|real})_{ij} \simeq-\frac{1}{2M}\left[\nu_i \nu_j
      +\frac{v_d(Y_{\nu_i}\nu_j+Y_{\nu_j}\nu_i)}{3\lambda}
      +\frac{Y_{\nu_i}Y_{\nu_j}v_d^2}{9\lambda^2 }\right].
\eea
We can also see that for $v_d \rightarrow 0$ (i.e. 
$\tan \beta=\frac{v_u}{v_d}\to \infty$) 
one obtains
\begin{eqnarray}
(m_{eff|real})_{ij} \simeq -\frac{\nu_i \nu_j}{2M}.
\label{Gaugino see-saw}
\end{eqnarray}
Note that this result can actually be obtained if $\nu_i >> \frac{Y_{\nu_i} v_d}{3 \lambda}$, and that this relation can be fulfilled with $v_d \sim v_u \sim 174$ GeV for suitable values of 
$\lambda$.
It means that decoupling right-handed neutrinos/singlinos and 
Higgsinos, the seesaw mechanism is generated through the mixing of 
left-handed neutrinos with gauginos. This is a characteristic feature 
of the seesaw in the well-known bilinear R-parity violation model (BRpV) \cite{Cita44paperHuitu}.

The seesaw in the $\mu\nu$SSM comes, in general, from the interplay of the above two 
limits. Namely, the limit where we suppress only certain Higgsino and 
gaugino mixing. Hence, taking $v_d \rightarrow 0$ in 
Eq. (\ref{Formula analitica real reescrita}),
which means quite pure gauginos but Higgsinos mixed
with right-handed neutrinos, 
we obtain
\begin{eqnarray}
(m_{eff|real})_{ij}\simeq   \frac{v_u^2}{6 \kappa \nu^c}Y_{\nu_i}Y_{\nu_j}
         (1-3 \delta_{ij})-\frac{1}{2 \, M_{\text{eff}}}\nu_i \nu_j\ ,
\label{Limit quasi no mixing Higgsinos gauginos}
\end{eqnarray}
As above, we remark that actually this result can be obtained 
if $\nu_i >>\frac{Y_{\nu_i} v_d}{3 \lambda}$.
The effective mass
$M_{\text{eff}} = M \left(1-\frac{v^4}{12 \kappa M  \nu^{c^3}  }\right)$
represents the mixing between gauginos
and Higgsinos-$\nu^c$ that is not completely suppressed 
in this limit.
Expression (\ref{Limit quasi no mixing Higgsinos gauginos}) is more general than the other two limits studied above.
On the other hand, for typical values of the parameters involved 
in the seesaw, $M_{\text{eff}}\approx M$, and therefore 
we get a simple formula that can be used to understand the seesaw
mechanism in this model in an qualitative way, that is
\begin{eqnarray}
(m_{eff|real})_{ij}\simeq \frac{v_u^2}
{6 \kappa \nu^c}Y_{\nu_i}Y_{\nu_j}
                   (1-3 \delta_{ij})-\frac{1}{2M} \nu_i \nu_j.
\label{Limit no mixing Higgsinos gauginos}
\end{eqnarray}
The simplicity of Eq. (\ref{Limit no mixing Higgsinos gauginos}), in contrast 
with the full formula given by Eq. (\ref{Formula analitica real reescrita}), 
comes from the fact that the mixing between gauginos and Higgsinos-$\nu^c$ is 
neglected. 

\begin{figure}[b]
 \begin{center}
\hspace*{-8mm}
    \epsfig{file=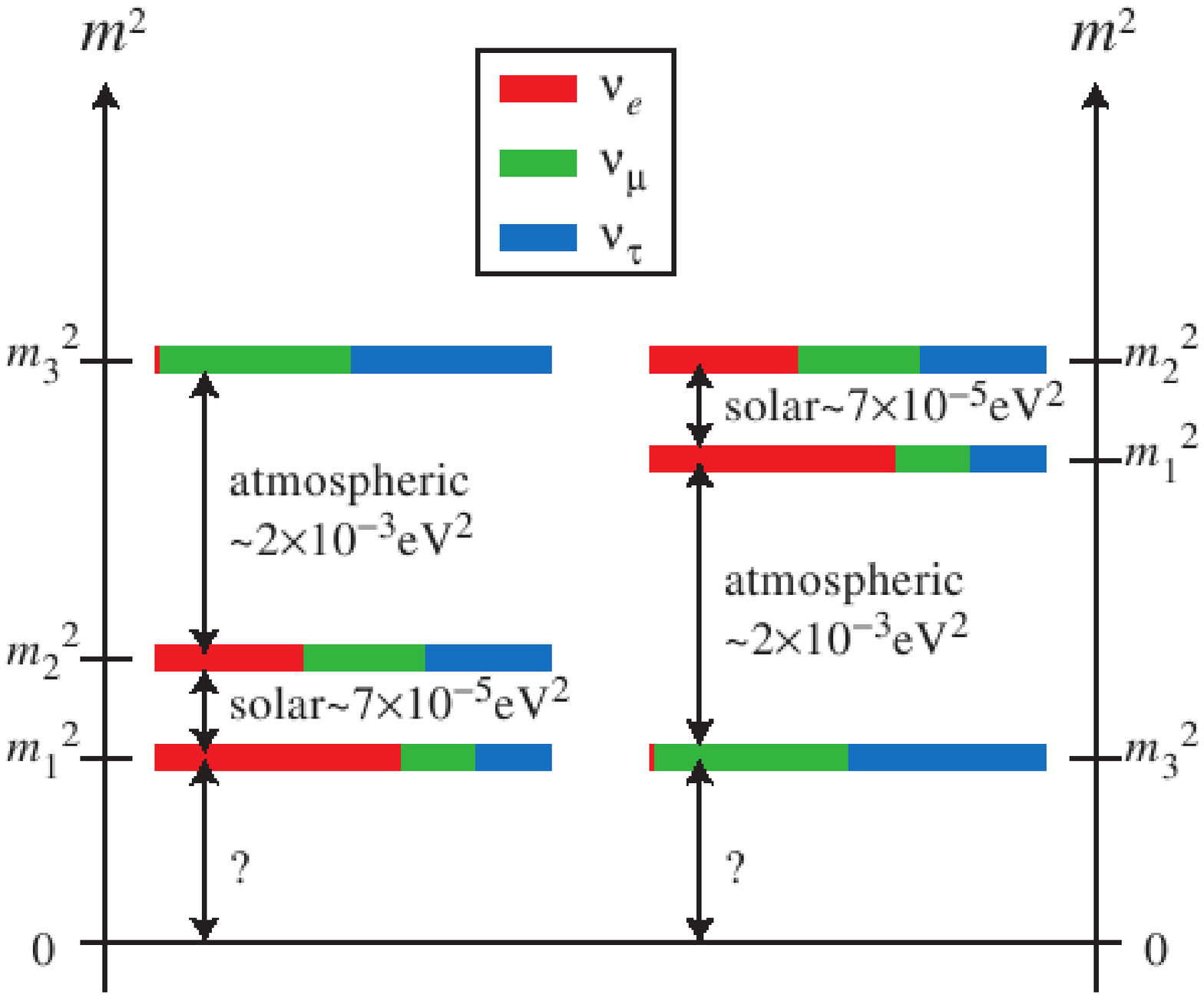 ,height=7.7cm,angle=0}
\captions{ The two possible hierarchies of neutrino masses as shown in
\cite{Dibujo Jerarquias}. The pattern on the left side corresponds to the 
normal hierarchy and is characterized by one heavy state with a very little 
electron neutrino component, and two almost degenerate light states with a 
mass difference which is the solar mass difference. 
The pattern on the right side corresponds to the inverted hierarchy and 
is characterized by two almost degenerate heavy states with a mass 
difference that is the solar mass difference, and a light state which 
has very little electron neutrino component. 
In both cases the mass difference between the heaviest/lightest eigenstate 
and the almost degenerate eigenstates is the atmospheric scale.}
    \label{ImagenJerarquias}
 \end{center}
\end{figure}

To continue the discussion of the seesaw in the
$\mu\nu$SSM, let us 
remind that two mass differences and mixing angles have been measured 
experimentally in the neutrino sector. The allowed $3 \sigma$ ranges for these parameters are shown in Table \ref{Synopsis}. 
We also show the compositions of the mass eigenstates in 
Fig. \ref{ImagenJerarquias} for the normal and inverted hierarchy cases.
For the discussion, hereafter we will use 
indistinctly the subindices $(1,2,3) \equiv (e,\mu,\tau)$.

\begin{table}[t]
\centering
\begin{tabular}{|c|c|c|c|c|}
\hline
$\Delta m_{sol}^2/10^{-5}\mathrm{\ eV}^2$ & $\sin^2\theta_{12}$ & 
            $\sin^2\theta_{13}$ & $\sin^2\theta_{23}$ &
            $\Delta m_{atm}^2/10^{-3}\mathrm{\ eV}^2$ \\[5pt]
\hline
7.14-8.19 & 0.263-0.375  & $<0.046$        & 
0.331-0.644  & 2.06-2.81 \\
\hline
\end{tabular}
\caption{\label{Synopsis} Allowed $3 \sigma$ ranges for the 
neutrino masses and mixings as discussed in \cite{ConstraintsFogli}.}
\end{table}

Due to the fact that the mass eigenstates have, in a good 
approximation, the same composition of $\nu_{\mu}$ and $\nu_{\tau}$ 
we start considering $Y_{\nu_2}=Y_{\nu_3}$ and 
$\nu_2=\nu_3$,  and therefore  
Eq. (\ref{Limit no mixing Higgsinos gauginos}) takes the form
\begin{align}
m_{eff}=\left(
\begin{array}{ccc}
d & c & c \\
c & A & B \\
c & B & A
\end{array}\right),
\label{meff toy model}
\end{align}

where
\begin{eqnarray}
d=-\frac{v_u^2}{3 \kappa \nu^c}Y_{\nu_1}^2-\frac{1}{2M}\nu_1^2, \nonumber \\
c=\frac{v_u^2}{6\kappa \nu^c}Y_{\nu_1}Y_{\nu_2}
                            -\frac{1}{2M}\nu_1 \nu_2, \nonumber \\
A=-\frac{v_u^2}{3 \kappa \nu^c}Y_{\nu_2}^2-\frac{1}{2M}\nu_2^2, \nonumber \\
B=\frac{v_u^2}{6\kappa \nu^c}Y_{\nu_2}^2-\frac{1}{2M}\nu_2^2.
\label{Parameters toy model}
\end{eqnarray}
The eigenvalues of this matrix are the following:
\begin{eqnarray}
\frac{1}{2}\left(A+B-\sqrt{8c^2+(A+B-d)^2}+d \right),
  \frac{1}{2}\left(A+B+\sqrt{8c^2+(A+B-d)^2}+d \right), 
  A-B\ ,                                    
\nonumber \\
\label{Eigenvalues toy model}
\end{eqnarray}
and the corresponding eigenvectors (for simplicity are not normalised) 
are
\begin{eqnarray}
& \left(-\frac{A+B+\sqrt{8c^2+(A+B-d)^2}-d}{2},c,c \right), \nonumber \\
& \left(\frac{-A-B+\sqrt{8c^2+(A+B-d)^2}+d}{2c},1,1 \right), \nonumber \\
& \left(0,-1,1 \right). 
\label{Eigenvectors toy model}
\end{eqnarray}
We have ordered the eigenvalues in such a way that it is clear how to 
obtain the normal hierarchy for the $\nu_{\mu}$-$\nu_{\tau}$ degenerate case. Then we see that $\sin^2\theta_{13}=0$ and $\sin^2\theta_{23}=\frac{1}{2}$, as 
in the tri-bimaximal mixing regime. Also we have enough freedom to fix the 
parameters in such a way that the experimental values for the mass 
differences and the remaining angle $\theta_{12}$ can be reproduced.
It is important to mention that the above two values of the angles 
are a consequence of considering the example with $\nu_{\mu}$-$\nu_{\tau}$ degeneration, and therefore valid even if we use the general 
formula (\ref{Formula analitica real reescrita}) instead of the simplified expression 
(\ref{Limit no mixing Higgsinos gauginos}). 
Notice that Eqs. (\ref{meff toy model}), (\ref{Eigenvalues toy model}) and (\ref{Eigenvectors toy model}) would be the same but with the corresponding
values of $A, B, c$ and $d$.

Let us remark that the fact that to obtain the correct neutrino angles is easy
in this kind of seesaw is due to the following characteristics: R-parity is broken and the relevant scale is the electroweak one. In a sense we are giving an answer
to the question why the mixing angles are so different in the quark and lepton sectors.

To show qualitatively how we can obtain an adequate seesaw with 
diagonal neutrino Yukawa couplings, let us first consider the limit\footnote{Actually this limit can be obtained 
taking $Y_{\nu_1} \rightarrow 0$, $\nu_1 \rightarrow 0$, implying $c \rightarrow 0$, and also $d \rightarrow 0$, and leading to similar conclusions. This limit means that the electron neutrino is decoupled from the other two neutrinos, having a negligible mass.} $c \rightarrow 0$ . In this limit the electron neutrino is the lightest neutrino, and is
completely decoupled from the rest.
The second eigenvector has no $\nu_e$ composition 
($\sin\theta_{12}\to 0$),
and it is half $\nu_{\mu}$ and half $\nu_{\tau}$. Understanding this case we can easy generalized the situation to the case 
$\sin\theta_{12}\neq 0$, switching on the parameter $c$. 
The eigenvalues in this limit are
\begin{eqnarray}
d \ , \ A+B, \ A-B,
\label{Eigenvalues toy model c to 0}
\end{eqnarray}
where
\bea
|d|=\left|\frac{v_u^2}{3 \kappa \nu^c}Y_{\nu_1}^2+\frac{1}{2M}\nu_1^2 \right|,  \nonumber \\
|A+B|=\left|\frac{v_u^2}{6 \kappa \nu^c}Y_{\nu_2}^2+\frac{1}{M}\nu_2^2 \right|, \nonumber \\
|A-B|= \frac{v_u^2}{2 \kappa \nu^c}Y_{\nu_2}^2. \nonumber \\
\eea
We can see that 
$\Delta m_{atm}^2 \sim |4AB|=\left|4(\frac{v_u^4 Y_{\nu_2}^4}{18 \kappa^2 \nu^{c^2}}-\frac{1}{4 M^2}\nu_2^4-\frac{v_u^2 Y_{\nu_2}^2 \nu_2^2}{12 M \kappa \nu^c})\right|$ 
and  $\Delta m_{sol}^2\sim |(A+B)^2-d^2|=\left|(\frac{v_u^2}{6 \kappa \nu^c}Y_{\nu_2}^2+\frac{1}{M}\nu_2^2)^2-(\frac{v_u^2}{3 \kappa \nu^c}Y_{\nu_1}^2+\frac{1}{2 M}\nu_1^2)^2\right|$.

It is important to note that we need $|A-B|>|A+B|$ for the normal hierarchy 
case, otherwise the $\theta_{12}$ angle is zero even when $c$ is not 
neglected. This is easy to obtain for $M>>2 \kappa \nu^c$. 
If $M \sim 2 \kappa \nu^c$,
using different signs for the effective Majorana and  
gaugino masses helps to fulfill the above
inequality.
For this to hold with our convention, one must take $M<0$.

In the inverted hierarchy scenario $|A-B|>|A+B|$ leads the 
angle $\theta_{12}$ to zero also with $c\neq 0$ which is not  
phenomenologically viable. Then we impose $|A-B|<|A+B|$.
Note that when $c$ is switched on, the parameter $d$ has to be large enough for having the associated neutrino with an intermediate mass, as corresponds to the inverted hierarchy scenario.
Therefore in this case we can also have easily the tri-bimaximal
mixing regime for $M << 2 \kappa\nu^c$. When 
$M \sim 2 \kappa \nu^c$, having $M>0$ helps to fulfill the
above condition.

Let us finally remark that we can get the complete tri-bimaximal mixing regime 
$\sin^2\theta_{13}= 0$, $\sin^2\theta_{23}=1/2$ and $\sin^2\theta_{12}=1/3$ fixing in Eq. (\ref{meff toy model}) $c=A+B-d$. In this way we obtain the eigenvalues
\begin{eqnarray}
-(A+B)+2d \ , \ 2(A+B)-d, \ A-B,
\label{Eigenvalues toy model c to 0ppp}
\end{eqnarray}
and from Eq. (\ref{Eigenvectors toy model}), after normalization, we arrive to
$\sin^2\theta_{12}=1/3$.

Breaking the degeneracy between the $Y_{\nu}$ and $\nu$ of the muon and 
tau neutrinos, it is possible to find more general solutions in the normal and inverted hierarchy cases. 
We will show this with numerical examples in the next section,
working always in the case $M \sim 2 \kappa \nu^c$.
Note also that in the case of degenerate $\nu_{\mu}$-$\nu_{\tau}$ parameters, 
as the Dirac CP phase always appears in the MNS matrix in the form $\sin\theta_{13}e^{i \delta}$ 
(see eq. 
(\ref{StandardParametrizationV}) below),
the SCPV effect is suppressed 
since $\sin\theta_{13}$ 
is negligible. This is not the case if we break the degeneration 
between $\nu_{\mu}$ and $\nu_{\tau}$.

When the vacuum is non CP-conserving the situation is more complicated 
since new relative phases are present, but the idea still holds.
In the next section we will use the above results to find 
numerical examples in the general case where also phases are generated 
through complex vacua. Examples where changing the sign of $M$  
the second and third eigenvalue are interchanged and the behavior is 
similar to the one described in this section.

\section{Results \label{section:results}}
In Section \ref{section:vacuua} we have given a simple argument 
to show that the $\mu\nu$SSM can violate CP spontaneously. 
In Section 3 we have discussed how to obtain correct neutrino masses and mixing angles.
In this section we sketch the numerical method used for the  
search of global minima of the $\mu\nu$SSM with SCPV,
giving rise also to an effective neutrino mass matrix 
that reproduces correctly the phenomenology of the neutrino sector 
according to observations. We also give some examples.

For simplicity, we assume that all the parameters appearing in the 
potential are diagonal in flavor space at the electroweak scale, except 
the trilinear $(A_\kappa \kappa)_{ijk}$ terms whose entries different from $(A_\kappa \kappa)_{iii}$
are relevant to break CP spontaneously. We introduce the following notation 
for the flavor diagonal free parameters of the scalar potential: 
$\kappa_i$, 
$Y_{\nu_i}$, $(A_\nu Y_\nu)_i$, $m^2_{\tilde L_i}$, 
$m^2_{\tilde \nu_i^c}$ with $i=1,2,3$ being flavor indices. 
Under this assumption, the neutral scalar potential in
(\ref{finalpotential})
is obviously simplified, and as a consequence also the minimization conditions (\ref{8vd}-\ref{7n}) are simplified.
In addition to the complex VEVs, the potential depends 
on $\lambda_i$, $\kappa_i$, $Y_{\nu_i}$, 
$(A_{\kappa} \kappa)_{ijk}$, $(A_{\lambda} \lambda)_i$, $(A_{\nu} Y_{\nu})_i$, $m_{H_d}$, $m_{H_u}$, 
$m_{\tilde \nu_i^c}$ and $m_{\tilde L_i}$. 

The strategy followed to 
find minima of the model consists of solving the minimization equations 
in terms of the soft parameters that are linear in those equations.
More precisely, the three minimization equations 
(\ref{7n}), corresponding to $\frac{\partial V}{\partial \chi_i}=0$, are used to solve the values of
$(A_{\nu} Y_{\nu})_i$. 
Using this result, Eqs. (\ref{7r}) for $i=2,3$, corresponding to $\frac{\partial V}{\partial \varphi_{\nu_{2,3}}^c}=0$,
are then solved for $(A_{\lambda} \lambda)_{2,3}$. Repeating the procedure using the equation (\ref{7v}), $\frac{\partial V}{\partial \varphi_v}=0$,
one obtains $(A_{\lambda} \lambda)_1$. Finally, Eq. (\ref{7r}) for $i=1$ is used 
to get $(A_{\kappa} \kappa)_{111}$. The conditions with 
respect to the moduli of the VEVs 
(\ref{8vd}-\ref{8n}) are used to get the squared soft masses.
Once this is done, we ensure that the critical point found 
(i.e. with non-vanishing phases for the VEVs) is a global minimum through a 
numerical procedure. As discussed in 
\cite{MuNuSSM2}, one has to check in particular that the minimum found
is deeper than the local minima with some or all the VEVs vanishing.

To accomplish the numerical task of finding global minima we 
need as inputs the eight moduli and seven phases of the VEVs, 
the $\lambda_i$, $\kappa_i$ and $Y_{\nu_i}$ couplings and the 
soft-trilinear terms $(A_{\kappa} \kappa)_{ijk}$ with $(i,j,k) \neq (1,1,1)$.
For simplicity, we assume a special structure for the latter: 
$(A_{\kappa} \kappa)_{222}=(A_{\kappa} \kappa)_{333}$, a common value for $(A_{\kappa} \kappa)_{ijk}$ 
with $i,j,k \neq 1$, and another common value for $(A_{\kappa} \kappa)_{ijk}$ with one or two indices equal 
to $1$. Moreover, let us recall that the modulus of the SUSY Higgs VEVs, can be determined from
$\textit{v}^2=v_d^2+v_u^2+\sum_i \nu_i^2 
\approx v_d^2+v_u^2$
with 
$\textit{v} \approx 174 \ GeV$, and the value of 
$\tan \beta=\frac{v_u}{v_d}$.

Once we find global minima, the next step is to build the neutralino 
mass matrix and to diagonalize it perturbatively in order to extract 
the effective neutrino mass matrix. Diagonalizing the effective neutrino mass matrix, we can extract the mass differences and the mixing angles 
of the neutrino sector and compare them with the data. 
The key for obtaining a phenomenologically viable neutrino sector, 
once we are in a global minimum, consists of varying either the neutrino Yukawa couplings, the left-handed sneutrino VEVs or the soft gaugino masses. 
This approach does not alter the vacuum structure previously obtained.

Let us now describe the details on how we proceed with the phenomenological 
analysis of the neutrino sector of the model. First, we assume for simplicity the GUT 
inspired relation between the gaugino masses $M_1$ and $M_2$, 
$M_1=\frac{\alpha_1^2}{\alpha_2^2} M_2$, implying $M_2 \simeq 2 M_1$ at low energy. 
As discussed in Section 3, one has to diagonalize 
the
neutrino effective mass matrix, $m_{eff}=-m^{T} \cdot M^{-1} \cdot m$.
Since it is a complex symmetric matrix, it can be diagonalized 
with an unitary transformation, as it is shown in Eqs. (\ref{nudagnudiag}) 
and (\ref{nudiag}). For the MNS matrix we follow the standard 
parameterization
\begin{eqnarray}\label{StandardParametrizationU}
 U_{MNS} & = &
    \diag(e^{\I\delta_{e}},e^{\I\delta_{\mu}},e^{\I\delta_{\tau}}) \cdot V \cdot 
    \diag(e^{-\I\phi_1/2},e^{-\I\phi_2/2},1),
\end{eqnarray}
where $\phi_1$ and $\phi_2$ are the Majorana phases 
and $V$ is given by 
\begin{align}
V=\left(\begin{array}{ccc}
c_{12}c_{13} & s_{12}c_{13} & s_{13} e^{-i \delta} \\
-c_{23}s_{12}-s_{23}s_{13}c_{12}e^{i \delta} & 
c_{23}c_{12}-s_{23}s_{13}s_{12}e^{i \delta} & s_{23}c_{13} \\
s_{23}s_{12}-c_{23}s_{13}c_{12}e^{i \delta} & 
-s_{23}c_{12}-c_{23}s_{13}s_{12} e^{i \delta} & c_{23}c_{13}
\label{StandardParametrizationV}
\end{array}\right).
\end{align}
Here $c_{ij} \equiv \cos\theta_{ij}$ and $s_{ij} \equiv \sin\theta_{ij}$ 
whereas  $\delta$ is the Dirac CP violating phase. 
The conventions used for extracting the mixing angles and the Majorana 
and Dirac phases from Eqs. (\ref{StandardParametrizationU}) and 
(\ref{StandardParametrizationV}) are outlined in Ref. \cite{Antusch:2003kp}.

Taking all the above into account,
we show in Table \ref{Parametros del minimo global} the parameters that 
characterize an example of a global minimum that breaks CP spontaneously.
The values of the soft parameters not determined by the minimization equations 
have been chosen to be $(A_{\kappa}\kappa)_{iii}=280$ GeV for 
$i \neq 1$, $(A_{\kappa}\kappa)_{ijk}=-40$ GeV for $i,j,k \neq 1$, and 
$(A_{\kappa}\kappa)_{ijk}=-120$ GeV for one or two indices equal to 1. 
In Table \ref{Punto jerarquia directa M1 negativa} 
we show the neutrino/neutralino inputs used in order 
to obtain a $\nu_{\mu}$-$\nu_{\tau}$ degenerated case with normal hierarchy,
producing values of masses and angles within the ranges of Table \ref{Synopsis}.
In particular, we obtain
$\sin^2 \theta_{13}\sim 0$ and
$\sin^2 \theta_{23}=0.5$,
as expected from the discussion in Section 3,
$\sin^2 \theta_{12}=0.323$, 
and neutrino masses 
$m_1=0.00305$ eV, $m_2=0.00949$ eV and $m_3=0.05091$ eV, 
producing $\Delta m_{solar}^2=8.08 \times 10^{-5}$ eV$^2$ and $\Delta m_{atm}^2=2.50 \times 10^{-3}$ eV$^2$.
The corresponding values of the soft terms calculated with the minimization equations are presented in Table 
\ref{Soft terms calculados del minimo global}.

It is worth noticing
that for this solution, the soft masses of the left-handed sneutrinos, 
$m_{\tilde L_i}$, do not need to be very different, and, actually, in this case they are almost degenerate $\sim 3700$ GeV. This can be understood using the minimization equations (\ref{8n}), neglecting the terms with products of Yukawas. When $\frac{Y_{\nu_i}}{\nu_i}=\frac{Y_{\nu_j}}{\nu_j}, \ \forall \ i,j$, one obtains $m^2_{\tilde L_i}=m^2_{\tilde L_j}$.
\begin{table}[t]
$$
\begin{array}{|c|c|c|}
\hline
\ \lambda_i=0.13 \ & \ \kappa_i=0.55 \ & \ \nu_i^c=1000 \ \textit{GeV}  
\\
\hline
\ \tan \beta=29 \ & \ \varphi_v=- \pi \ & \ \varphi_{\nu_1^c}=\frac{\pi}{7} \ \\
\hline
\ \varphi_{\nu_2^c}=\varphi_{\nu_3^c}=-\frac{\pi}{7} \ & \ \chi_1=-\frac{\pi}{6} \ & \ \chi_2=\chi_3=\frac{\pi}{6} \ \\
\hline
\end{array}
$$
\caption{Numerical values of the relevant input parameters for a global minimum that 
breaks CP spontaneously.}
\label{Parametros del minimo global}
\end{table}
\begin{table}[t]
$$
\begin{array}{|c|c|c|}
\hline
\ Y_{\nu_1}=4.25 \times 10^{-7} \ & \ Y_{\nu_2}=Y_{\nu_3}=1.36 \times 10^{-6} & \ M_1=-340 \ \textrm{GeV} \ \\
\hline
\ \nu_1=3.88 \times 10^{-5} \ \textrm{GeV} \ & \ \nu_2=\nu_3=1.24 \times 10^{-4} \ \textrm{GeV}  \\
\cline{1-2}
\end{array}
$$
\caption{Numerical values of the neutrino/neutralino inputs 
that reproduce 
the neutrino experimental constraints, and correspond to the normal hierarchy 
scenario.
}
\label{Punto jerarquia directa M1 negativa}
\end{table}
\begin{table}[t]
$$
\begin{array}{|c|c|c|}
\hline
\ (A_{\nu} Y_{\nu})_1 \simeq -0.0031 \ \textrm{GeV} \ & \ (A_{\nu} Y_{\nu})_2 \simeq -0.010 \ \textrm{GeV} \ & \ (A_{\nu} Y_{\nu})_3 \simeq -0.010 \ \textrm{GeV} \ \\
\hline
\ (A_{\lambda} \lambda)_1 \simeq -1487 \ \textrm{GeV} \ & \ (A_{\lambda} \lambda)_2 \simeq -679 \ \textrm{GeV} \ & \ (A_{\lambda} \lambda)_3 \simeq -679 \ \textrm{GeV} \ \\
\hline
\ (A_{\kappa} \kappa)_{111} \simeq -0.25 \ \textrm{GeV} \ & \ m^2_{H_d} \simeq 7.0325 \times 10^7 \ \textrm{GeV}^2 \ & \ m^2_{H_u} \simeq -47200 \ \textrm{GeV}^2 \ \\
\hline
\ m^2_{\tilde \nu_1^c} \simeq 260140 \ \textrm{GeV}^2 \ & \ m^2_{\tilde \nu_2^c} \simeq -100820 \ \textrm{GeV}^2 \ & \ m^2_{\tilde \nu_3^c} \simeq -100820 \ \textrm{GeV}^2 \ \\
\hline
\ m_{\tilde L_1}^2\simeq \ m_{\tilde L_2}^2=m_{\tilde L_3}^2=1.37 \times 10^{7} \ \textrm{GeV}^2  \\
\cline{1-1}
\end{array}
$$
\caption{Values of the soft terms calculated with the minimization 
equations for the global minimum associated to the parameters shown in 
Table \ref{Parametros del minimo global}. 
}
\label{Soft terms calculados del minimo global}
\end{table}
However, we have to point out that the values obtained for other soft parameters are not so natural in a SUSY framework.
Notice for example that $A_\nu\sim -7$ TeV, $A_{\lambda_1}\sim -11$ TeV,  whereas
$A_{{\kappa}_{111}}\sim -0.5$ GeV. Indeed,
this is a consequence of the particular solution shown in Table \ref{Parametros del minimo global}.

Although it is non-trivial to find realistic solutions, since
many minima which apparently are acceptable, at the end of the day turn out to be
false minima, 
we have been able to find more sensible solutions. This is the case of the one shown in
Table \ref{Parametros del minimo global2}, with the values of the
input soft parameters
$(A_{\kappa} \kappa)_{iii}=-150$ GeV for $i \neq 1$, 
$(A_{\kappa} \kappa)_{ijk}=75$ GeV for $i,j,k \neq 1$ and 
$(A_{\kappa} \kappa)_{ijk}=-50$ GeV for one or two indices equal to $1$.
For example, 
lowering the values of $\nu^c$ one is able to lower the trilinear terms
$A_\nu\sim -3$ TeV in order to fulfill Eqs. (\ref{7n}) (also lowering 
$\kappa$ contributes to this result), and also the soft masses  
$m_{\tilde L_i}\sim 2.8$ TeV, as shown in Table 
\ref{Soft terms calculados del minimo global2}. 
Lowering $\lambda$ one is able to lower the trilinears
$A_{\lambda_1}\sim -1.5$ TeV, $A_{\lambda_{2,3}}\sim -840$ GeV, in order
to fulfill Eqs. (\ref{7v}) and (\ref{7r}).
Notice finally that the use of non-degenerate $\nu^c_i$ allows
to increase the trilinear  
$A_{{\kappa}_{111}}\sim 36$ GeV.
In Table \ref{Punto jerarquia directa M1 negativa2} 
we show the corresponding neutrino/neutralino inputs 
producing values of masses and angles within the ranges of Table \ref{Synopsis}.

\begin{table}[t]
$$
\begin{array}{|c|c|c|}
\hline
\ \lambda_i=0.10 
 & \ \kappa_i=0.35 
& \ \nu_1^c=835 \ \textit{GeV} \ , \ \nu_2^c=\nu_3^c=685 \ \textit{GeV}  \ \\
\hline
\ \tan \beta=29 \ & \ \varphi_v=- \pi \ & \ \varphi_{\nu_1^c}=\frac{\pi}{7} \ \\
\hline
\ \varphi_{\nu_2^c}=\varphi_{\nu_3^c}=-\frac{\pi}{7} \ & \ \chi_1=-\frac{\pi}{6} \ & \ \chi_2=\chi_3=\frac{\pi}{6} \ \\
\hline
\end{array}
$$
\caption{Numerical values of the relevant inputs for the second global minimum discussed in the text, that breaks CP spontaneously.}
\label{Parametros del minimo global2}
\end{table}

\begin{table}[t]
$$
\begin{array}{|c|c|c|}
\hline
\ Y_{\nu_1}=5.4 \times 10^{-7} \ & \ Y_{\nu_2}=Y_{\nu_3}=9.2 \times 10^{-7} & \ M_1=-340 \ \textrm{GeV} \ \\
\hline
\ \nu_1=3.7 \times 10^{-5} \ \textrm{GeV} \ & \ \nu_2=\nu_3=8.8 \times 10^{-5} \ \textrm{GeV}  \\
\cline{1-2}
\end{array}
$$
\caption{Numerical values of the neutrino/neutralino inputs for the second
global minimum discussed in the text,
that reproduce 
the neutrino experimental constraints and correspond to the normal hierarchy 
scenario.
}
\label{Punto jerarquia directa M1 negativa2}
\end{table}
\begin{table}[t]
$$
\begin{array}{|c|c|c|}
\hline
\ (A_{\nu} Y_{\nu})_1 \simeq -0.00209 \ \textrm{GeV} \ & \ (A_{\nu} Y_{\nu})_2 \simeq -0.00294 \ \textrm{GeV} \ & \ (A_{\nu} Y_{\nu})_3 \simeq -0.00294 \ \textrm{GeV} \ \\
\hline
\ (A_{\lambda} \lambda)_1 \simeq -156 \ \textrm{GeV} \ & \ (A_{\lambda} \lambda)_2 \simeq -84 \ \textrm{GeV} \ & \ (A_{\lambda} \lambda)_3 \simeq -84 \ \textrm{GeV} \ \\
\hline
\ (A_{\kappa} \kappa)_{111} \simeq 12.7 \ \textrm{GeV} \ & \ m^2_{H_d} \simeq 5.36 \times 10^6 \ \textrm{GeV}^2 \ & \ m^2_{H_u} \simeq -37910 \ \textrm{GeV}^2 \ \\
\hline
\ m^2_{\tilde \nu_1^c} \simeq 51035 \ \textrm{GeV}^2 \ & \ m^2_{\tilde \nu_2^c} \simeq 69155 \ \textrm{GeV}^2 \ & \ m^2_{\tilde \nu_3^c} \simeq 69155 \ \textrm{GeV}^2 \ \\
\hline
\ m_{\tilde L_1}^2=8.07 \times 10^6 \ \textrm{GeV}^2 \ & \ \ m_{\tilde L_2}^2=3.92 \times 10^6 \ \textrm{GeV}^2 \ & \ m_{\tilde L_3}^2=3.92 \times 10^{6} \ \textrm{GeV}^2  \\
\hline
\end{array}
$$
\caption{Values of the soft terms calculated with the minimization 
equations for the second global minimum discussed in the text, associated to the parameters shown in Table~\ref{Parametros del minimo global2}. 
}
\label{Soft terms calculados del minimo global2}
\end{table}

Modifying the values of the angles we can also obtain other interesting solutions.
See for example the one shown in Tables \ref{Parametros del minimo global333}, \ref{Punto jerarquia directa M1 negativa333}, and \ref{Soft terms calculados del minimo global333}.
In this case the values of the input soft parameters are chosen to be $(A_\kappa \kappa)_{iii}=-200$ GeV for $i \neq 1$, $(A_\kappa \kappa)_{ijk}=125$ GeV for $i,j,k \neq 1$ and $(A_\kappa \kappa)_{ijk}=-75$ GeV for one or two indices equal to $1$.
Notice that now the values obtained for the soft terms
are also of this order. In particular, the trilinears are $A_{\nu_1}\sim -657$ GeV, $A_{\nu_{2,3}}\sim -429$ GeV,
$A_{\lambda_1}\sim -990$ GeV, $A_{\lambda_{2,3}}\sim -830$ GeV, and
$A_{\kappa_{111}}\sim 100$ GeV. For the soft masses we obtain
$m_{\tilde L_1}\sim 628$ GeV, $m_{\tilde L_{2,3}}\sim 950$ GeV. 

A general analysis of the parameter space, finding other interesting complex vacua,
is obviously extremely complicated given the large number of parameters involved, and beyond the scope of this paper. Nevertheless, we have checked that
other sensible solutions can indeed be obtained modifying adequately the parameters.  
In the following we will work with the solution associated to the parameters of
Table~\ref{Parametros del minimo global}, since the discussion below is essentially valid for other solutions.
Our strategy will consist of varying the neutrino/neutralino inputs
$Y_{\nu_i}, \ \nu_i \ \textrm{and} \ M_1$ in such a way that the derived 
neutrino mass differences and mixing angles are within the
ranges of Table 1. 
As mentioned above, this procedure will not alter the vacuum structure found. 
Notice in this respect that gaugino masses do not contribute to the minimization equations, and that the values of $Y_{\nu_i}$ and $\nu_i$ are very small.
Let us also mention that this strategy can indeed be applied to the much more simple issue of analyzing real vacua.
In particular, it was shown in \cite{MuNuSSM2} that
many global minima with real VEVs can be found.
For them neutrino/neutralino inputs 
$Y_{\nu_i}, \ \nu_i, \ M_1$, similar to those studied here are also
valid.

\begin{table}[t]
$$
\begin{array}{|c|c|c|}
\hline
\ \lambda_i=0.10   \ & \ \kappa_i=0.42  \ & \ \nu_1^c=850 \ \textit{GeV} \ , \ \nu_2^c=\nu_3^c=550 \ \textit{GeV}  \ \\
\hline
\ \tan \beta=29 \ & \ \varphi_v=- \pi \ & \ \varphi_{\nu_1^c}=\frac{\pi}{5} \ \\
\hline
\ \varphi_{\nu_2^c}=\varphi_{\nu_3^c}=-\frac{\pi}{5} \ & \ \chi_1=-\frac{\pi}{3} \ & \ \chi_2=\chi_3=\frac{\pi}{3} \ \\
\hline
\end{array}
$$
\caption{Numerical values of the relevant inputs for the third global minimum 
discussed in the text, that breaks CP spontaneously.}
\label{Parametros del minimo global333}
\end{table}

\begin{table}[t]
$$
\begin{array}{|c|c|c|}
\hline
\ Y_{\nu_1}=1.9 \times 10^{-7} \ & \ Y_{\nu_2}=Y_{\nu_3}=8.5 \times 10^{-7} & \ M_1=-100 \ \textrm{GeV} \ \\
\hline
\ \nu_1=6 \times 10^{-5} \ \textrm{GeV} \ & \ \nu_2=\nu_3=4.9 \times 10^{-5} \ \textrm{GeV}  \\
\cline{1-2}
\end{array}
$$
\caption{Numerical values of the neutrino/neutralino inputs 
for the third global minimum discussed in the text, that reproduce 
the neutrino experimental constraints and correspond to the normal hierarchy 
scenario.
}
\label{Punto jerarquia directa M1 negativa333}
\end{table}
\begin{table}[t]
$$
\begin{array}{|c|c|c|}
\hline
\ (A_{\nu} Y_{\nu})_1 \simeq -0.000125 \ \textrm{GeV} \ & \ (A_{\nu} Y_{\nu})_2 \simeq -0.000365 \ \textrm{GeV} \ & \ (A_{\nu} Y_{\nu})_3 \simeq -0.000365 \ \textrm{GeV} \ \\
\hline
\ (A_{\lambda} \lambda)_1 \simeq -99 \ \textrm{GeV} \ & \ (A_{\lambda} \lambda)_2 \simeq -83 \ \textrm{GeV} \ & \ (A_{\lambda} \lambda)_3 \simeq -83 \ \textrm{GeV} \ \\
\hline
\ (A_{\kappa} \kappa)_{111} \simeq 41.9 \ \textrm{GeV} \ & \ m^2_{H_d} \simeq 3.6 \times 10^6 \ \textrm{GeV}^2 \ & \ m^2_{H_u} \simeq -25118 \ \textrm{GeV}^2 \ \\
\hline
\ m^2_{\tilde \nu_1^c} \simeq -24393 \ \textrm{GeV}^2 \ & \ m^2_{\tilde \nu_2^c} \simeq 208377 \ \textrm{GeV}^2 \ & \ m^2_{\tilde \nu_3^c} \simeq 208377 \ \textrm{GeV}^2 \ \\
\hline
\ m_{\tilde L_1}^2=394777 \ \textrm{GeV}^2 \ & \ \ m_{\tilde L_2}^2=903528 \ \textrm{GeV}^2 \ & \ m_{\tilde L_3}^2=903528 \ \textrm{GeV}^2  \\
\hline
\end{array}
$$
\caption{Values of the soft terms calculated with the minimization 
equations for the third global minimum discussed in the text, associated to the parameters shown in 
Table~\ref{Parametros del minimo global333}. 
}
\label{Soft terms calculados del minimo global333}
\end{table}


As noted in Sect. \ref{section:neutrino} we have chosen $M_1 < 0$ in order 
to guaranty a viable $\theta_{12}$ angle. 
It is worth pointing out here that a redefinition of the parameters leaving the Lagrangian invariant can be made, in such a way that 
$M_1$ becomes positive and other parameters such as the VEVs become negative, 
describing indeed the same physics. 
In our convention the VEVs,
$v_d$, $v_u$, $\nu_i^c$, $\nu_i$, are always taken positive.
 
We would also like to stress that all the numerical results have been obtained without 
any approximation, that is, with the exact expression of the $10 \times 10$ 
neutralino mass matrix, calculating numerically the effective neutrino mass 
matrix and diagonalizing it.

Let us first study how the neutrino mass differences depend on the inputs. 
In Sect. \ref{section:neutrino} we showed that in this 
scenario there are 
two different contributions to the seesaw mechanism; the one involving 
right-handed neutrinos (and Higgsinos) given by 
$\frac{(Y_{\nu_i} v_u)^2}{2 \kappa \nu^c}$, where the Dirac and Majorana 
masses are parameterized by $Y_{\nu_i} v_u$ and $2 \kappa \nu^c$, respectively, 
and the contribution coming from the gaugino seesaw given by 
$\frac{(g_1 \nu_i)^2}{M_1}+\frac{(g_2 \nu_i)^2}{M_2}$, where the 
Dirac and  Majorana masses are parameterized by $g_{\alpha} \nu_i$ and 
$M_{\alpha}$, respectively, with $\alpha=1,2$. 
\begin{figure}[t!]
 \begin{center}
\hspace*{-8mm}
    \begin{tabular}{cc}
    \epsfig{file=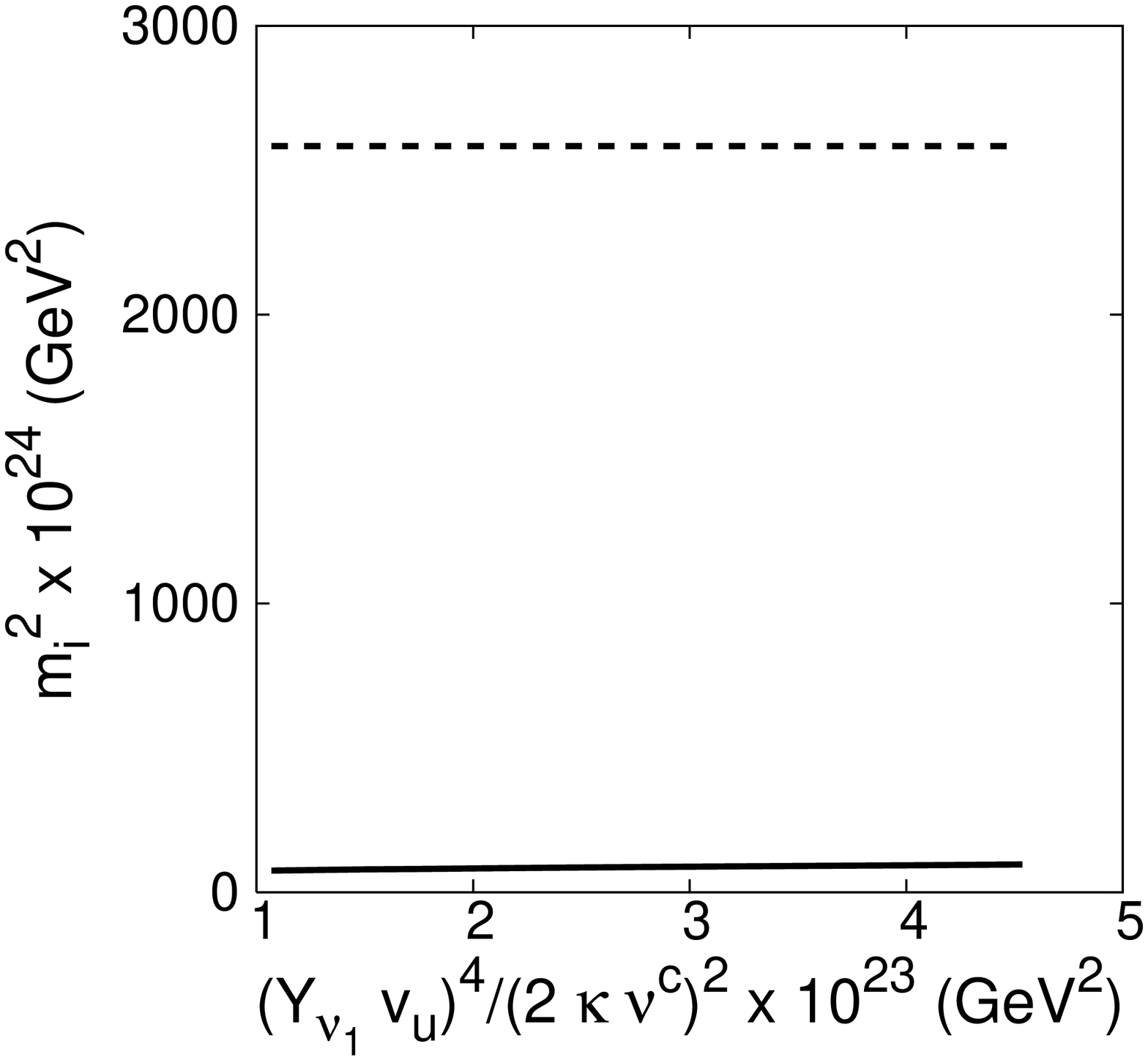 
,height=8.5cm,angle=-90}
    \hspace*{0mm}&\hspace*{-3mm}
      \epsfig{file=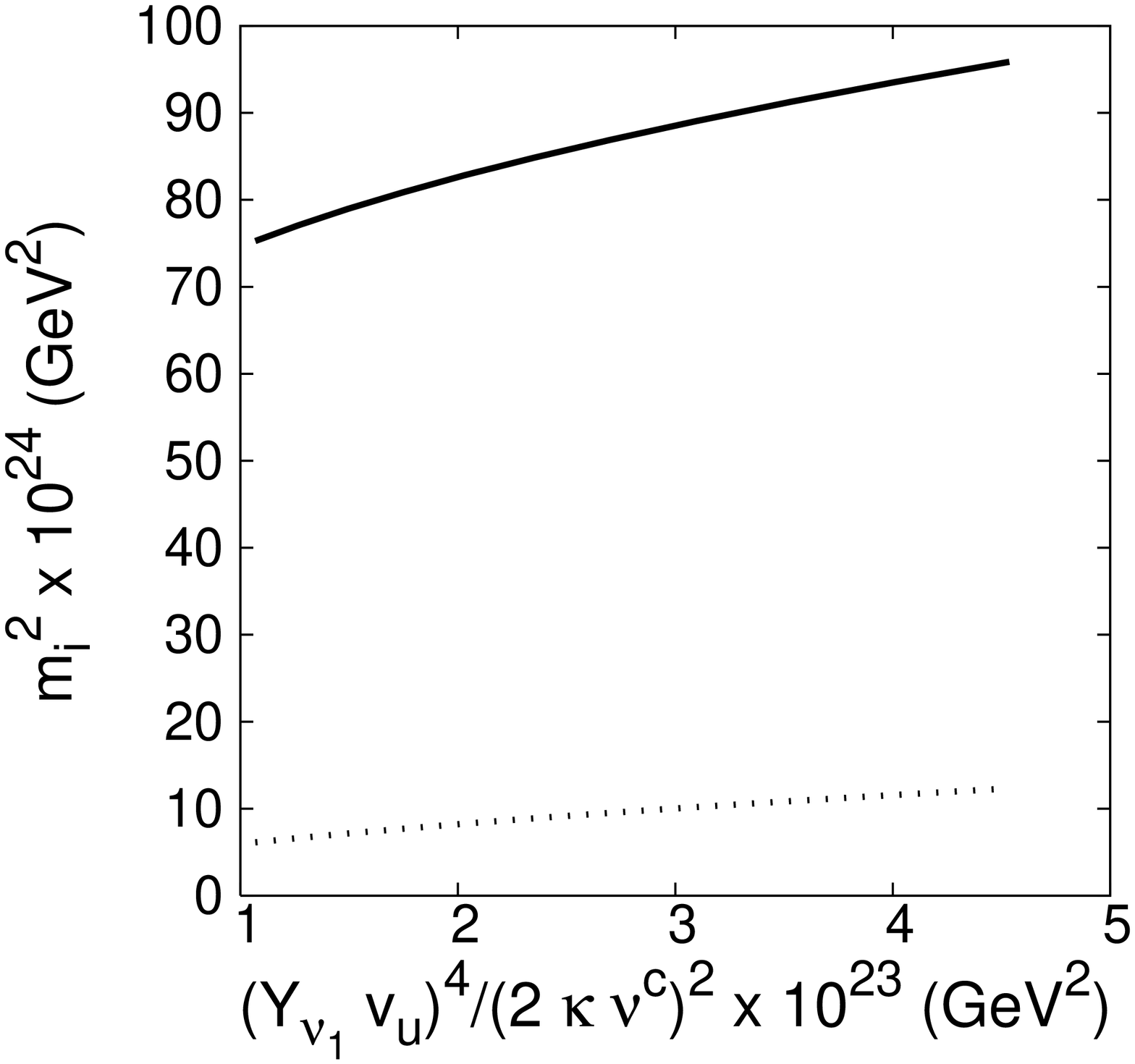,height=8.5cm,angle=-90}
      \\ & \\
      \\ \hspace*{-1.1cm} (a) & \hspace*{-1.1cm} (b)\\
    \epsfig{file=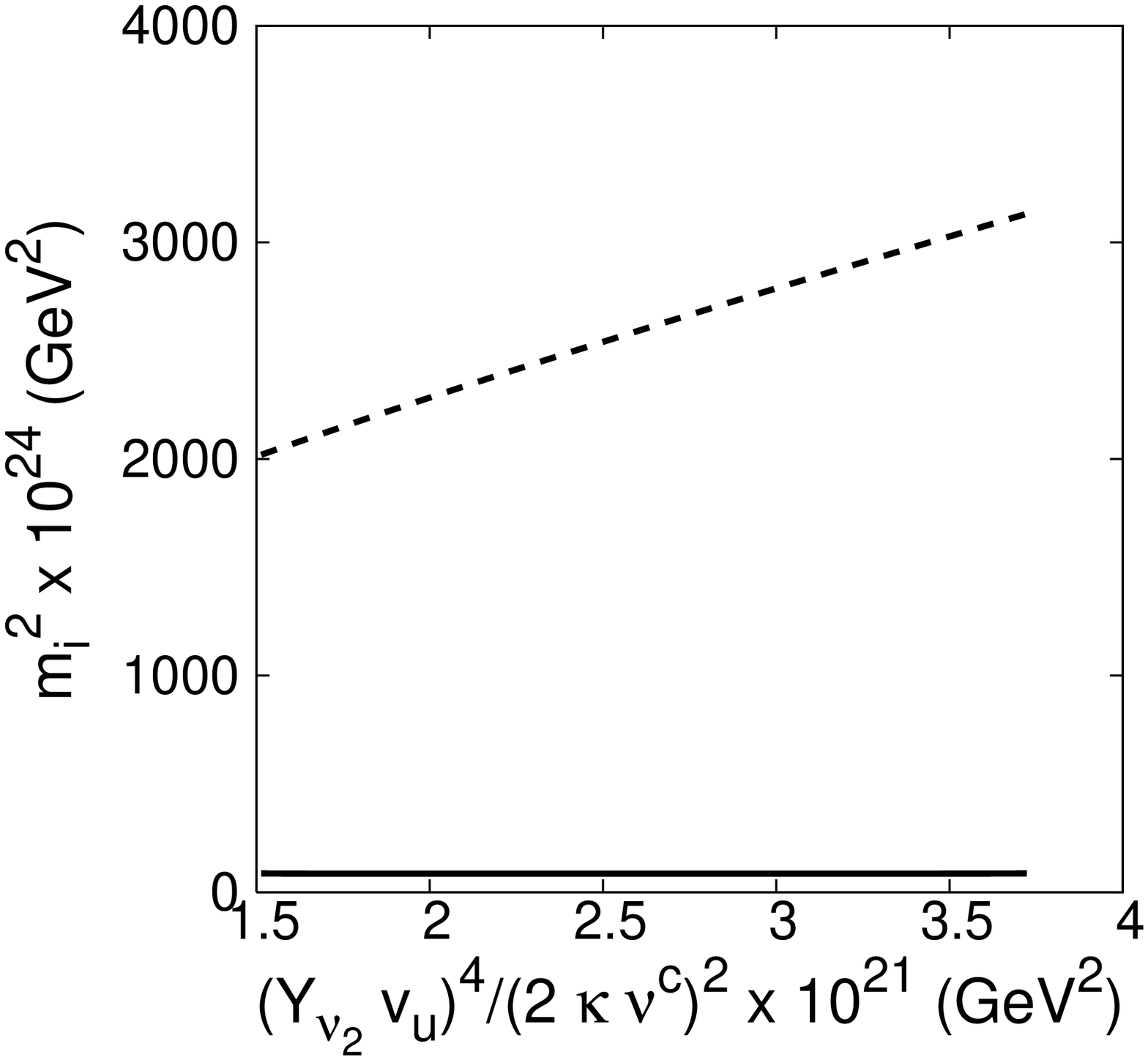 
,height=8.5cm,angle=-90}
    \hspace*{0mm}&\hspace*{-3mm}
      \epsfig{file=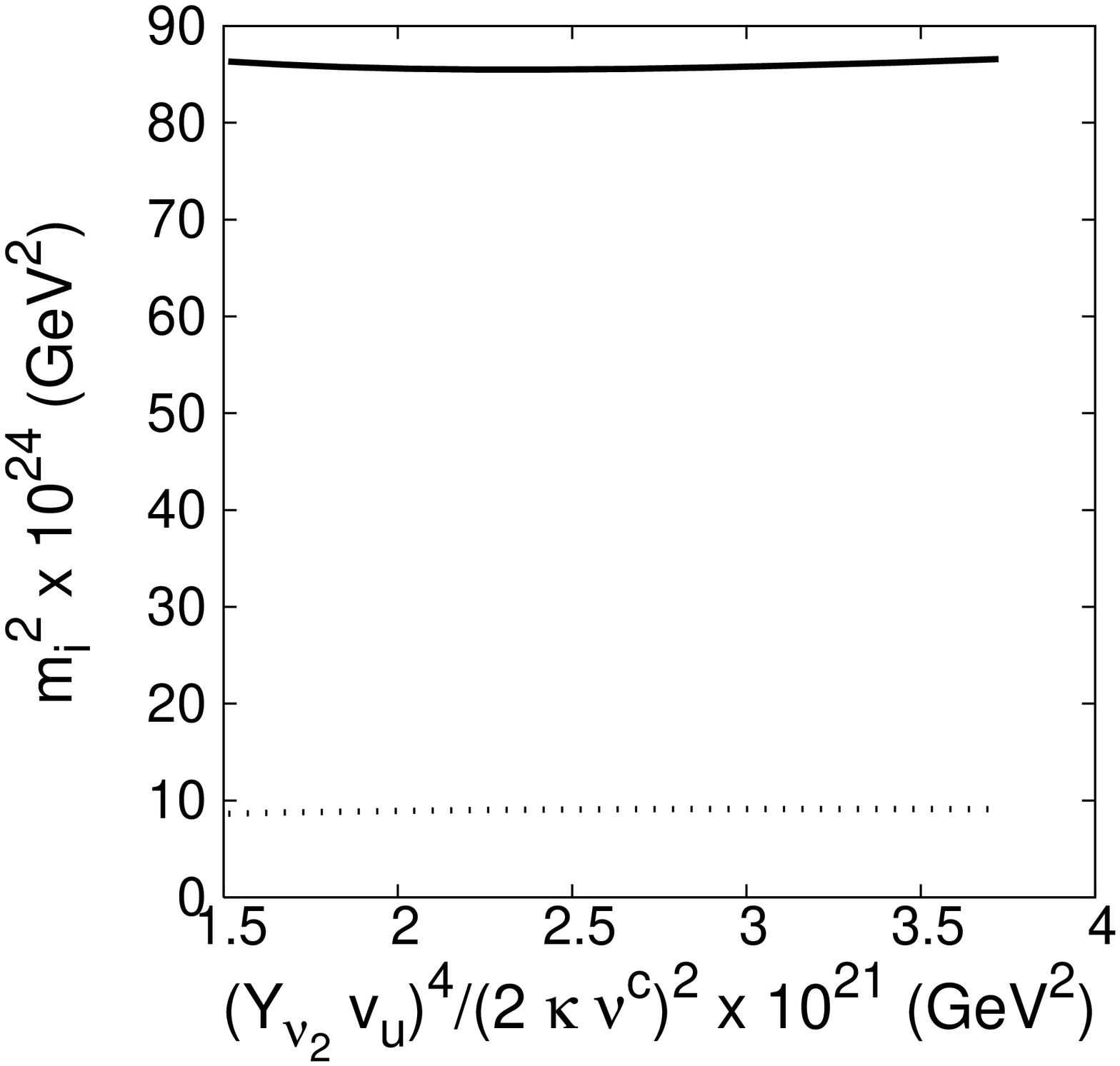,height=8.5cm,angle=-90}
      \\ & \\
      \\ \hspace*{-1.1cm} (c) & \hspace*{-1.1cm} (d)\\
    \end{tabular}
\captions{Squared neutrino masses versus 
$(Y_{\nu_i}v_u)^4/(2 \kappa \nu^c)^2$. 
(a) and (b) show for $i=1$ the two heaviest and lightest neutrinos, respectively. 
The same for 
(c) and (d) but for $i=2$.
}
    \label{Masas grandes y pequenas DsqOverMaj Indice}
 \end{center}
\end{figure}

\begin{figure}[t!]
 \begin{center}
\hspace*{-8mm}
    \begin{tabular}{cc}
    \epsfig{file=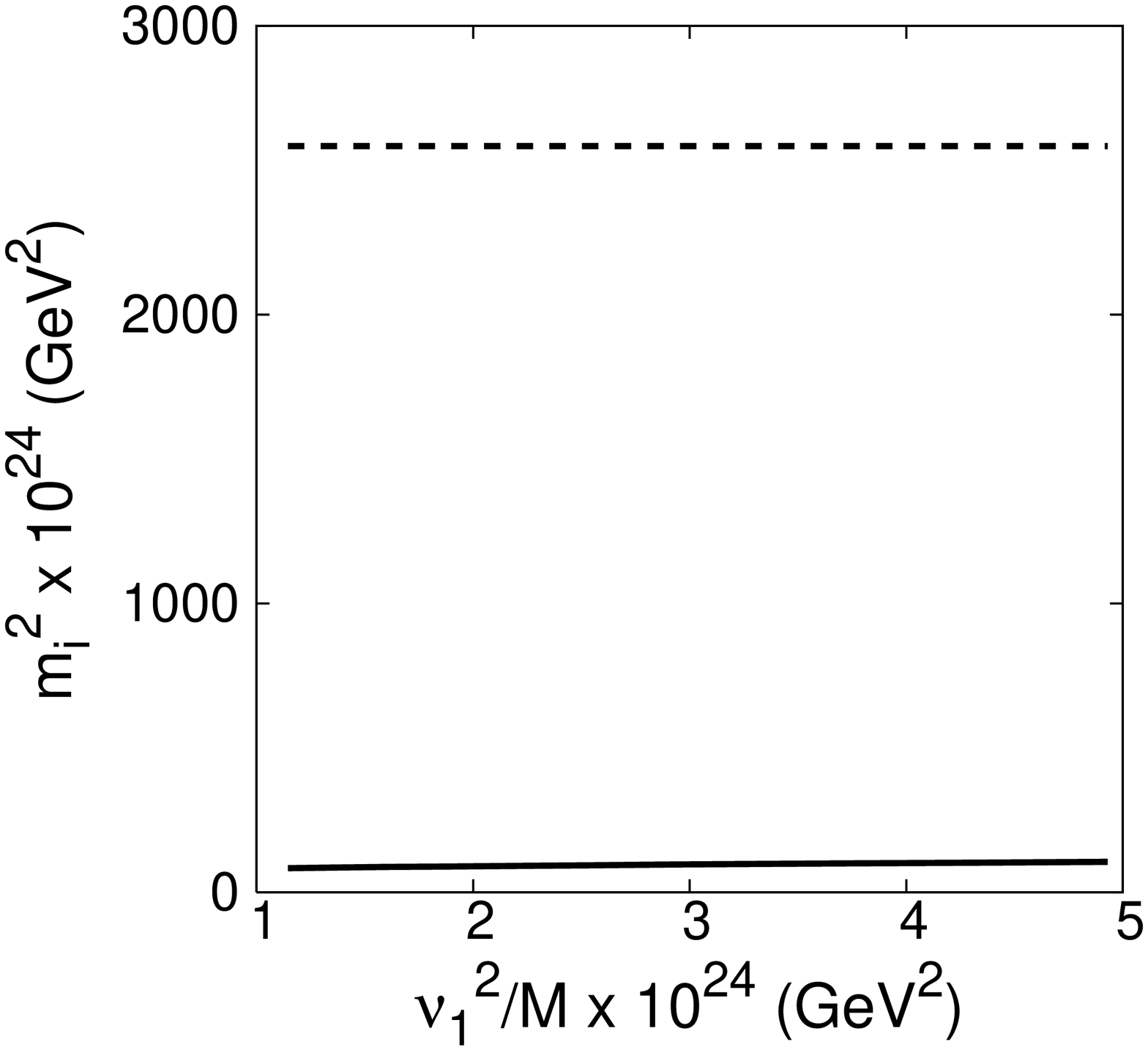 
,height=8.5cm,angle=-90}
    \hspace*{0mm}&\hspace*{-3mm}
      \epsfig{file=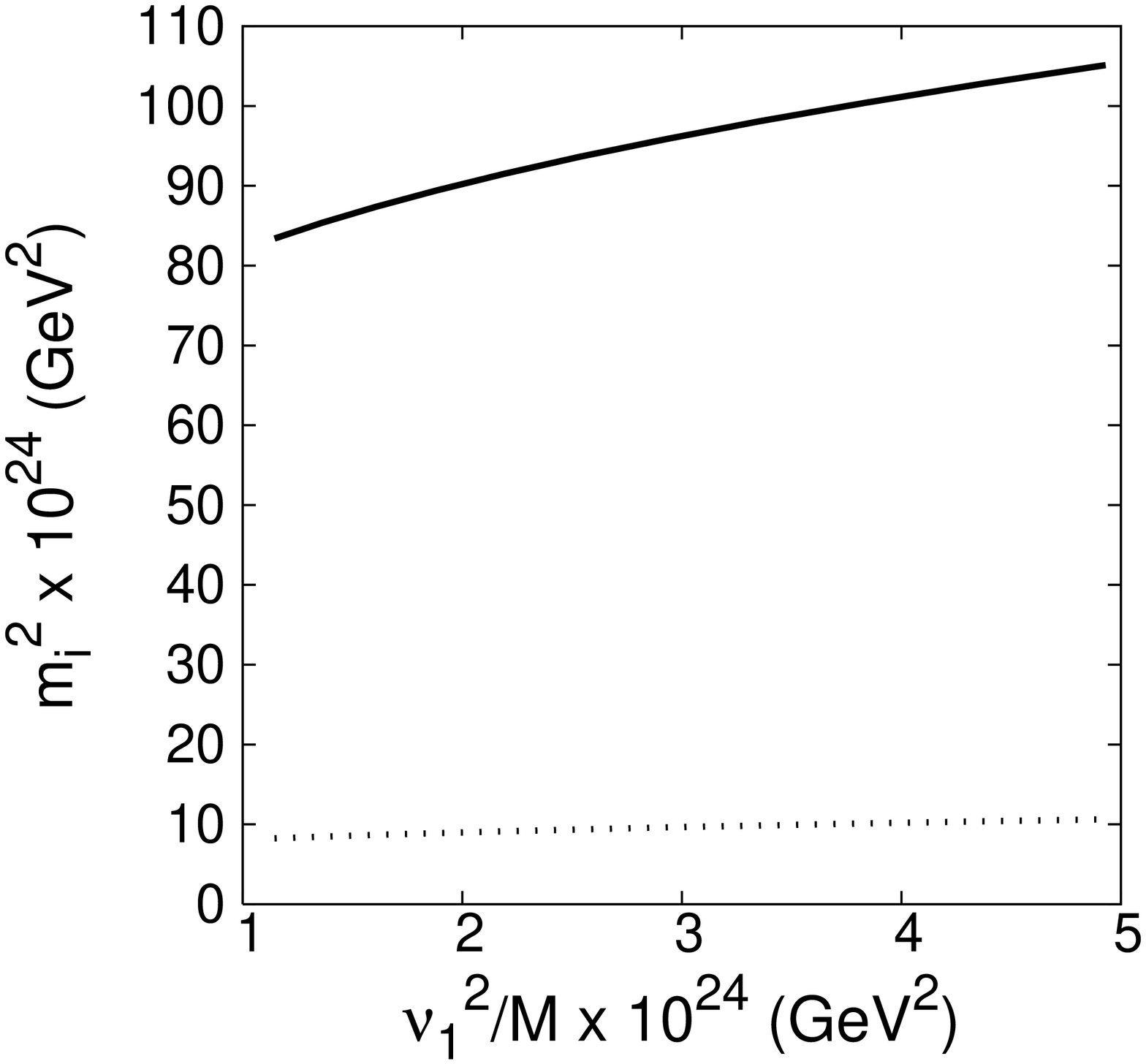,height=8.5cm,angle=-90}
      \\ & \\
      \\  \hspace*{-1.1cm} (a) & (b)\\
    \epsfig{file=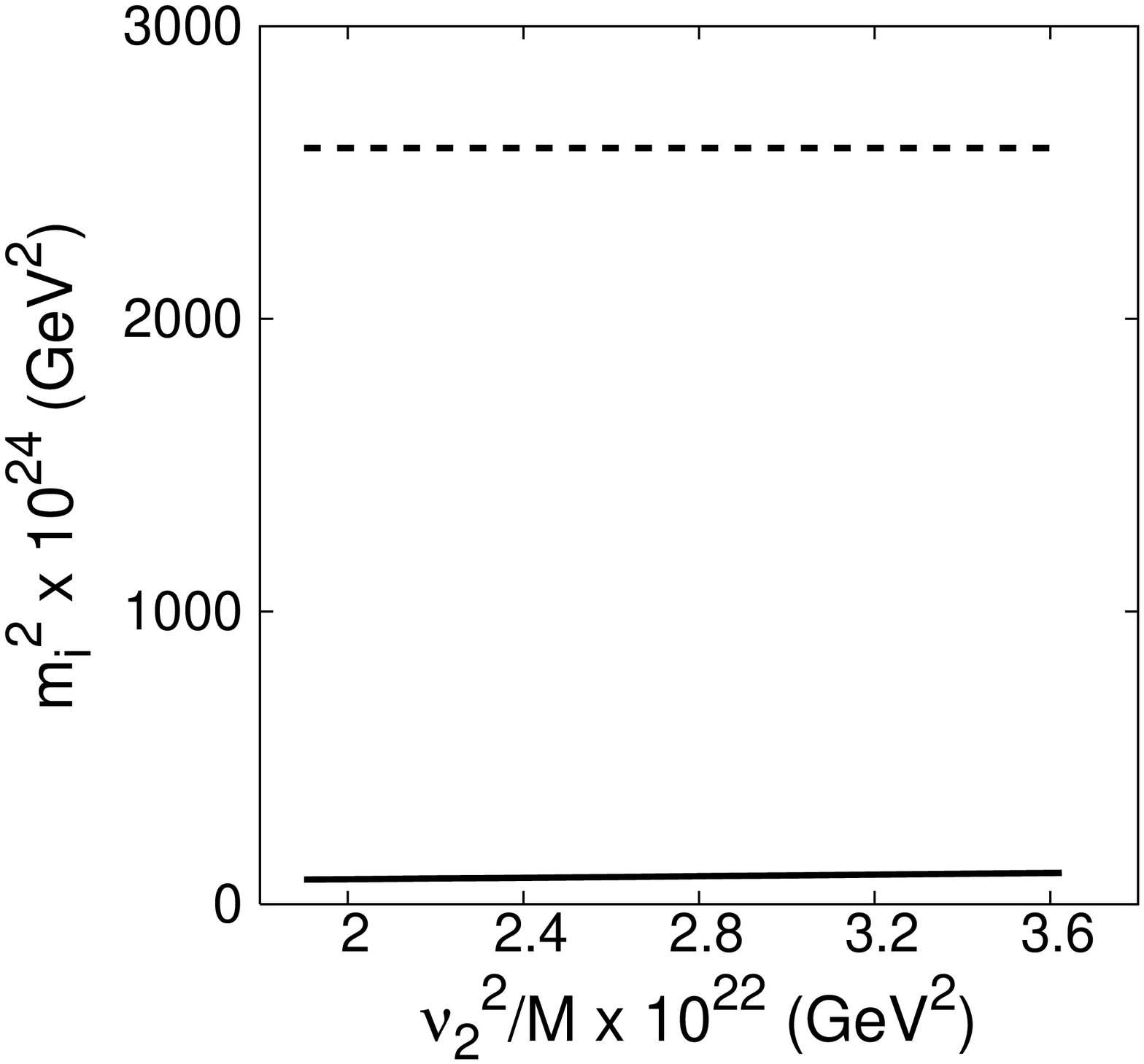 
,height=8.5cm,angle=-90}
    \hspace*{0mm}&\hspace*{-3mm}
      \epsfig{file=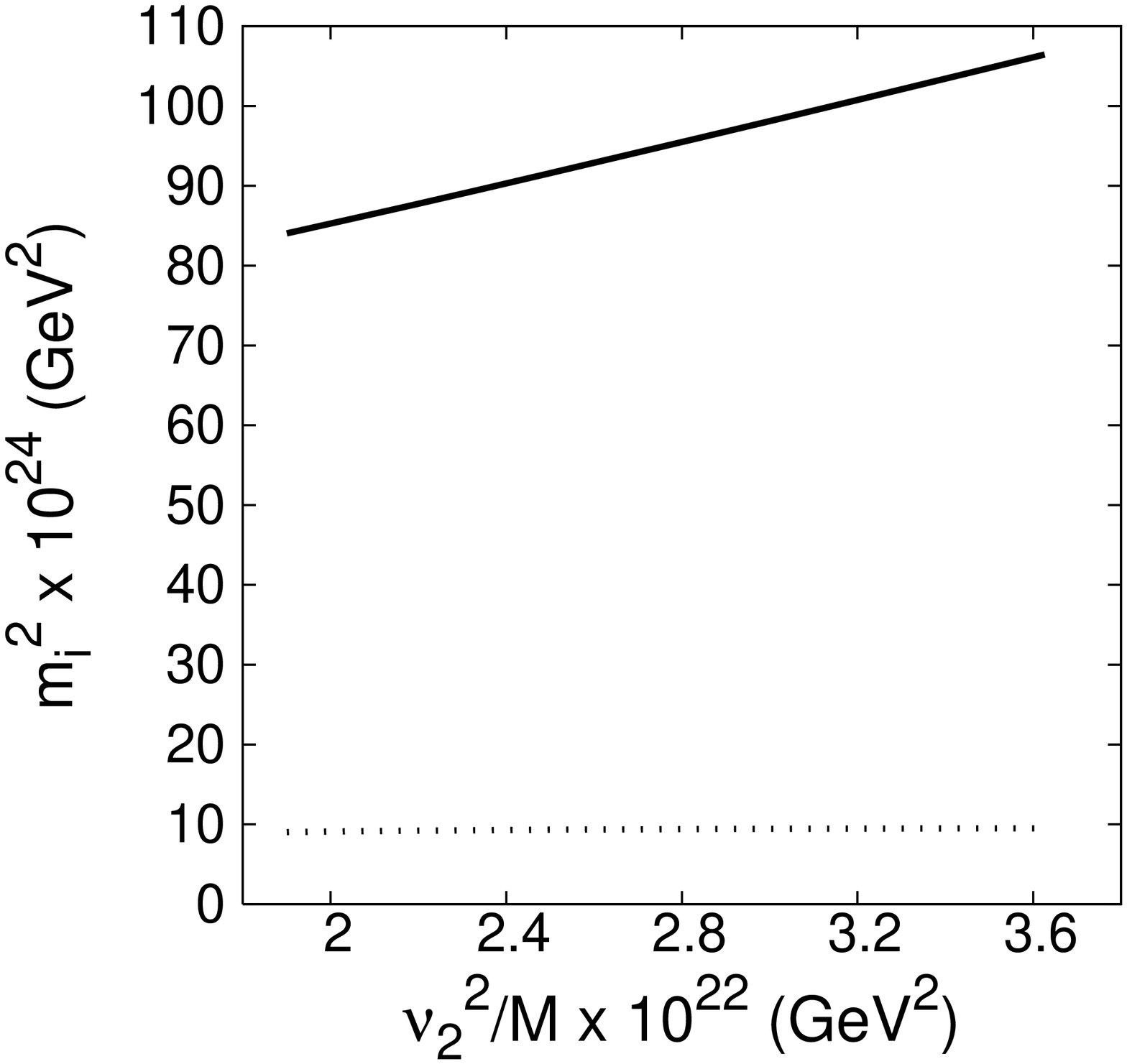,height=8.5cm,angle=-90}
      \\ & \\
      \\  \hspace*{-1.1cm} (c) &  \hspace*{-1.1cm} (d)\\
    \end{tabular}
\captions{
The same as in Fig. 2 but for 
the squared neutrino masses versus $[(g_1 \nu_i)^2/M_1+(g_2 \nu_i)^2/M_2]^2$.
}
    \label{Masas grandes y pequenas nu}
 \end{center}
\end{figure}

Figs. \ref{Masas grandes y pequenas DsqOverMaj Indice}a and \ref{Masas grandes y pequenas DsqOverMaj Indice}b show that the heaviest eigenvalue (dashed line) has very 
little electron-neutrino component, 
as expected in the normal hierarchy scenario (see Fig. \ref{ImagenJerarquias}),
and therefore it 
does not depend on $(Y_{\nu_1} v_u)^2/(2 \kappa \nu^c)$, 
whereas the intermediate (solid line) and lightest 
(dotted line)
eigenvalues, that have sizeable electron-neutrino components, grow with 
this term. As a consequence of the latter, the squared solar mass difference 
grows as well. 
On the other hand, following the arguments related to 
Eq. (\ref{Limit no mixing Higgsinos gauginos}), 
we can see in Figs. \ref{Masas grandes y pequenas DsqOverMaj Indice}c and \ref{Masas grandes y pequenas DsqOverMaj Indice}d 
that the heaviest eigenvalue 
is controlled by the 
contribution of the seesaw with right-handed neutrinos having an important 
muon/tau neutrino composition, thus we observe how the heaviest eigenvalue 
grows with $(Y_{\nu_2} v_u)^2/(2 \kappa \nu^c)$ and, as a consequence, the 
squared atmospheric mass difference 
grows accordingly. 
The variation with  $(Y_{\nu_3} v_u)^2/(2 \kappa \nu^c)$ 
is analogue. 

Fig. \ref{Masas grandes y pequenas nu} is analogous to Fig. 
\ref{Masas grandes y pequenas DsqOverMaj Indice} but showing  
the squared neutrino mass differences dependence on the gaugino seesaw component. 
In this case, because the heaviest eigenstate (dashed line) practically 
does not mix with the electron neutrino we can see that it does not vary with 
$((g_1 \nu_i)^2/M_1+(g_2 \nu_i)^2/M_2)^2$ for $i=1,2,3$.
On the other hand, the intermediate eigenstate grows with the mixing 
with the gauginos, as explained in Sect.  
\ref{section:neutrino} with $M_1<0$, 
therefore the squared solar mass 
difference also grows. 

\begin{figure}[t!]
 \begin{center}
\hspace*{1cm}
     \epsfig{file=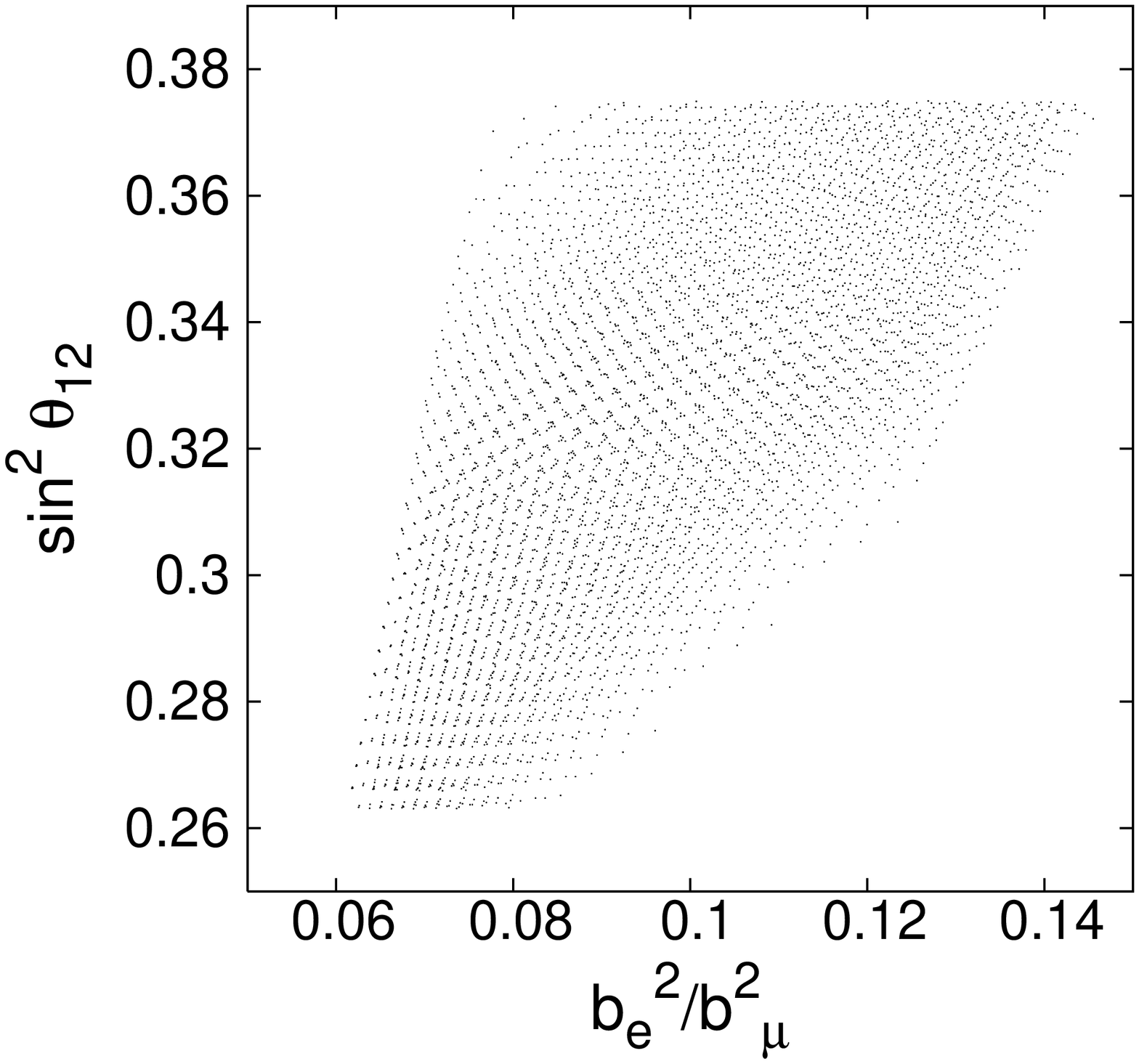 ,height=8.5cm,angle=-90}
\captions{Variation of the solar mixing angle with respect to the relevant term that 
controls its evolution, $b_e^2/b_{\mu}^2$. 
}
    \label{Senos1}
 \end{center}
\end{figure}

Let us now discuss the mixing angles. Note that in the 
$\nu_{\mu}$-$\nu_{\tau}$ degenerate case with normal hierarchy and $M_1<0$ we have obtained $\sin^2 \theta_{13} =0$ and $\sin^2 \theta_{23} = \frac{1}{2}$. In Fig. \ref{Senos1} 
we present the variation of $\sin^2 \theta_{12}$  with the ratio of the 
parameters that control the gaugino seesaw, $b_e^2/b_\mu^2$, where 
for the sake of simplicity we take $b_i=Y_{\nu_i} v_d+3 \lambda \nu_i$ and we do not 
consider the complicated factors containing phases in 
Eqs. (\ref{Parameters of the approximate analytical expression 5}).

\begin{figure}[t!]
 \begin{center}
\hspace*{-8mm}
    \begin{tabular}{cc}
      \epsfig{file=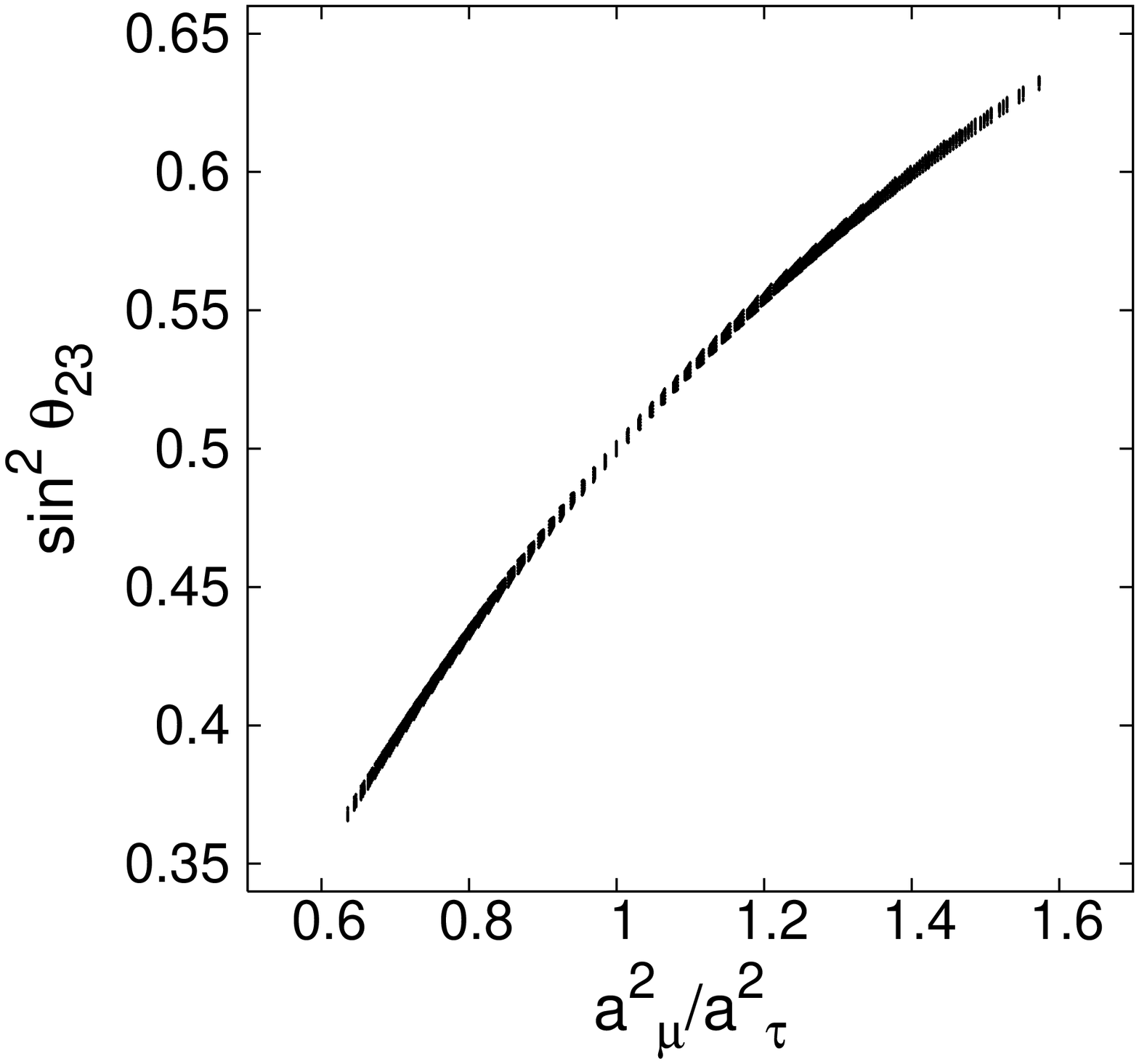,height=8.5cm,angle=-90}
        \hspace*{0mm}&\hspace*{-3mm}
\epsfig{file=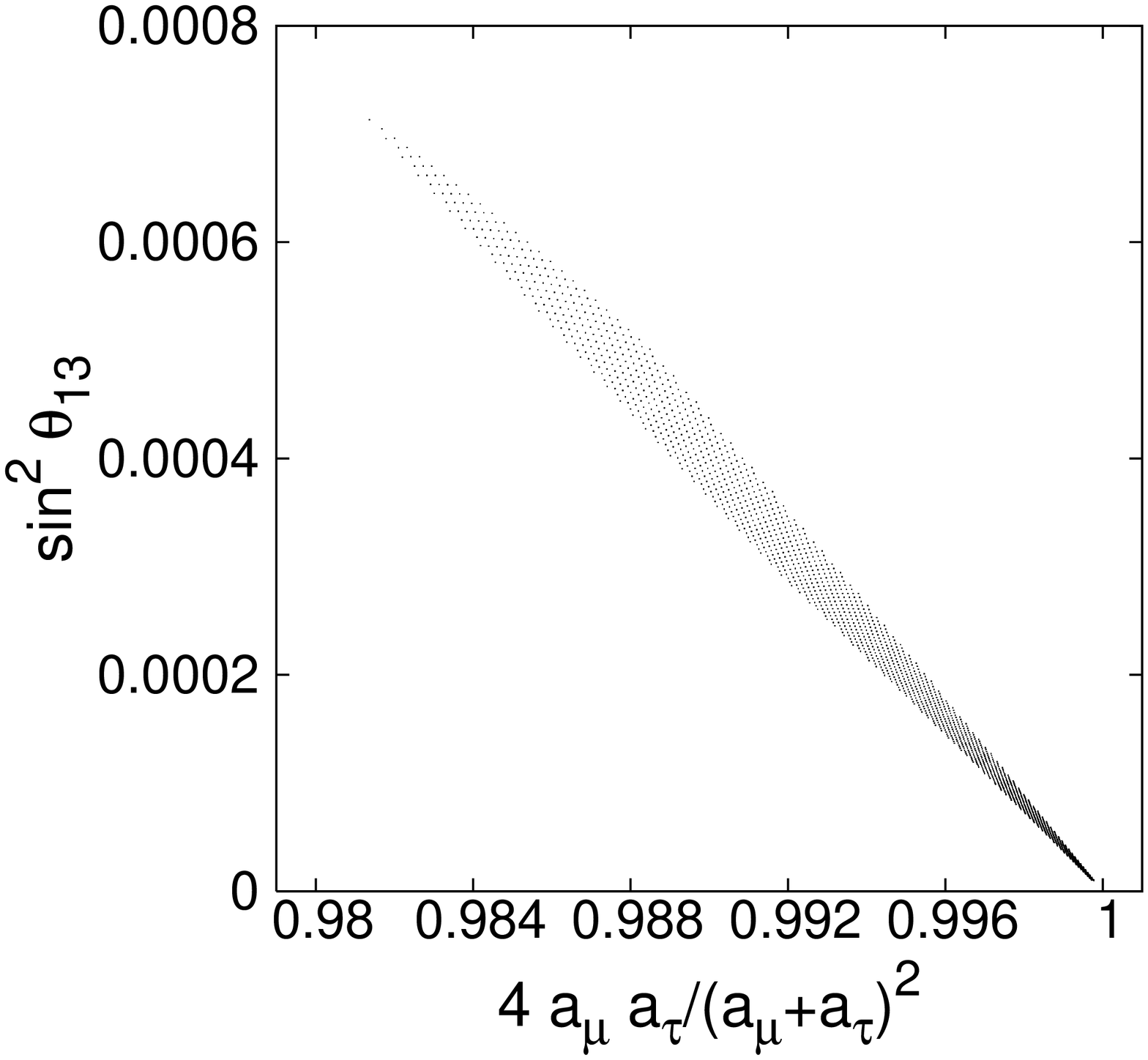,height=8.5cm,angle=-90}
      \\ & \\
       \hspace*{-1.1cm} (a) & \hspace*{-1cm} (b)
    \end{tabular}
\captions{
(a) The variation of $\sin^2 \theta_{23}$ with respect to the relevant 
term that controls its evolution, $a_{\mu}^2/a_{\tau}^2$. 
(b) The variation of $\sin^2 \theta_{13}$ with respect to the 
term that measures the $\nu_{\mu}$-$\nu_{\tau}$ degeneracy.}
    \label{Senos2}
 \end{center}
\end{figure}

To obtain
results different from $\sin^2 \theta_{23}\sim \frac{1}{2}$  
and $\sin^2 \theta_{13}\sim 0$, in the following we 
consider the possibility of having different values for 
the $Y_{\nu}$ and $\nu$ parameters for $\mu$ and $\tau$ neutrinos. 
We show in Fig. \ref{Senos2}a $\sin^2 \theta_{23}$ as a function of the 
ratio of the term that controls the Higgsino-$\nu^c$ seesaw, 
$a_{\mu}^2/a_{\tau}^2$.
When $a_{\mu}/a_{\tau}$ goes to $1$, the $\nu_{\mu}$-$\nu_{\tau}$ 
degeneracy is recovered and $\sin^2 \theta_{23}$ goes to $1/2$ as expected. 
In Fig. \ref{Senos2}b we show $\sin^2 \theta_{13}$ 
as a function of $\frac{4 a_{\mu} a_{\tau} }{(a_{\mu}+a_{\tau})^2}$ that is a 
good measure of the degeneration in this case. 
Note that when
$4a_{\mu}a_{\tau}/(a_{\mu}+a_{\tau})^2 \rightarrow 1$ the degeneracy 
is recovered and 
$\sin^2 \theta_{13} \rightarrow 0$ as expected.
The parameters $a_i$ have been defined 
in Eq. (\ref{Parameters of the approximate analytical expression}). 
Let us point out that $\sin^2 \theta_{13}<10^{-3}$ since we are breaking the 
degeneration between $\mu$ and $\tau$ neutrinos but 
the term that controls the higgsino-$\nu^c$ seesaw for the first 
family is very small compared to the other two families.

\begin{figure}[t!]
 \begin{center}
\hspace*{-8mm}
    \begin{tabular}{cc}
    \epsfig{file=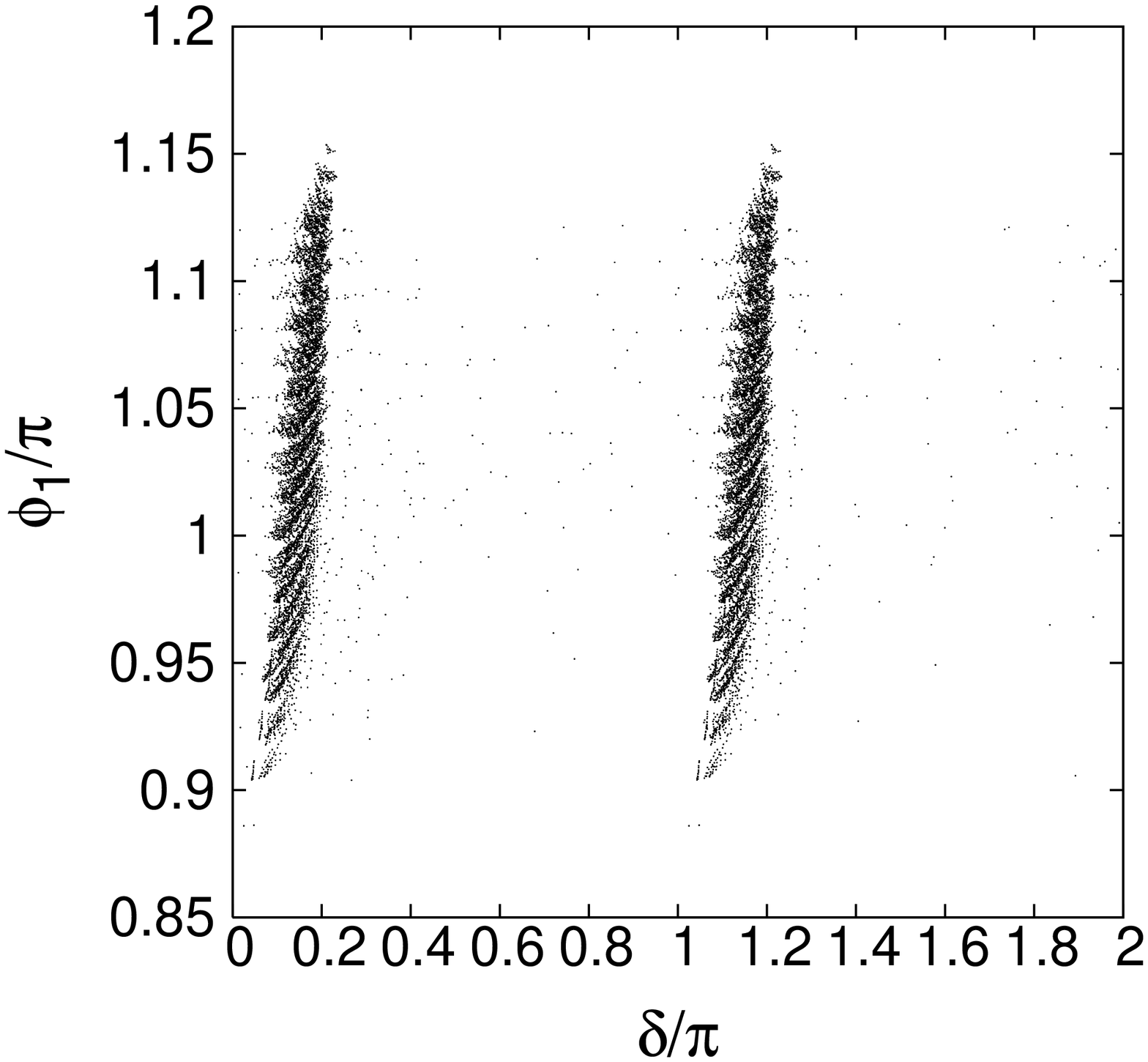
,height=8.5cm,angle=-90}
    \hspace*{0mm}&\hspace*{-3mm}
      \epsfig{file=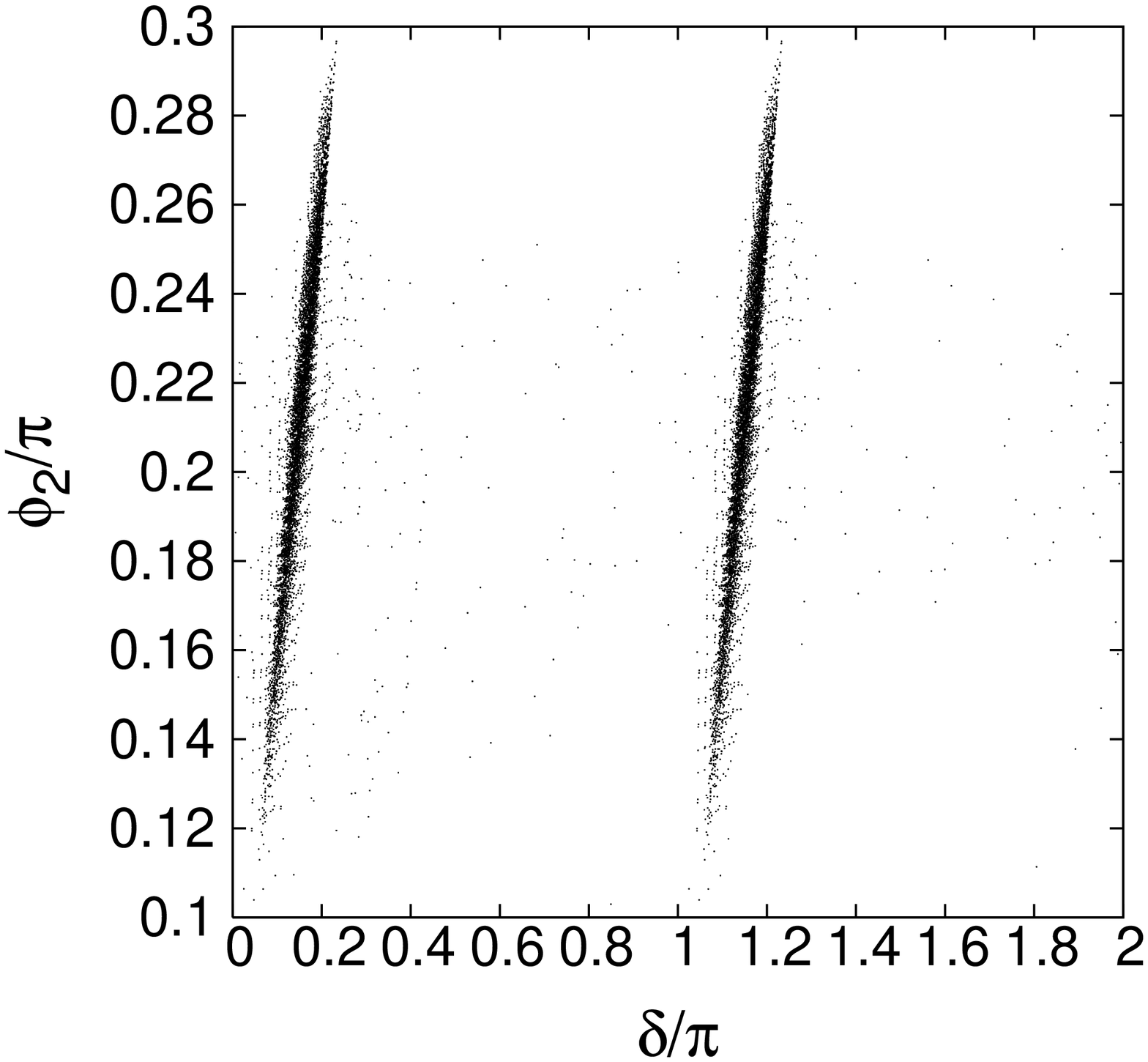,height=8.5cm,angle=-90}
      \\ & \\
      \hspace*{-1.1cm} (a) & \hspace*{-1cm} (b)
    \end{tabular}
\captions{
$\delta-\phi_1$ plane (a) and $\delta-\phi_2$ plane (b)
for the scenario with normal hierarchy and negative gaugino 
masses $M<0$,
varying simultaneously 
$Y_{\nu_i}, \nu_i, M_1$. 
}
    \label{Fases CP}
 \end{center}
\end{figure}

As mentioned previously, the $\mu\nu$SSM with SCPV also predicts non-zero 
CP phases in the MNS matrix. We have checked numerically that for each of the experimentally allowed regions found, 
the two Majorana CP phases and the Dirac CP phase are different from zero.
This fact is reflected in Fig. \ref{Fases CP} where we present two plots
in the $\delta-\phi_1$ and 
$\delta-\phi_2$ planes (Dirac-Majorana CP phases) constructed varying all the inputs in the 
neutrino sector. 
However, it is fair to say that
due to the smallness of $\sin^2 \theta_{13} \sim 10^{-3}$ 
in this region, the CP violation effects of the phases of the VEVs turn out to be suppressed 
in the MNS matrix because the Dirac CP phase always appears 
in the form $\sin \theta_{13} e^{i\delta}$.

\begin{figure}[t]
 \begin{center}
\hspace*{-8mm}
\begin{tabular}{cc}
\epsfig{file=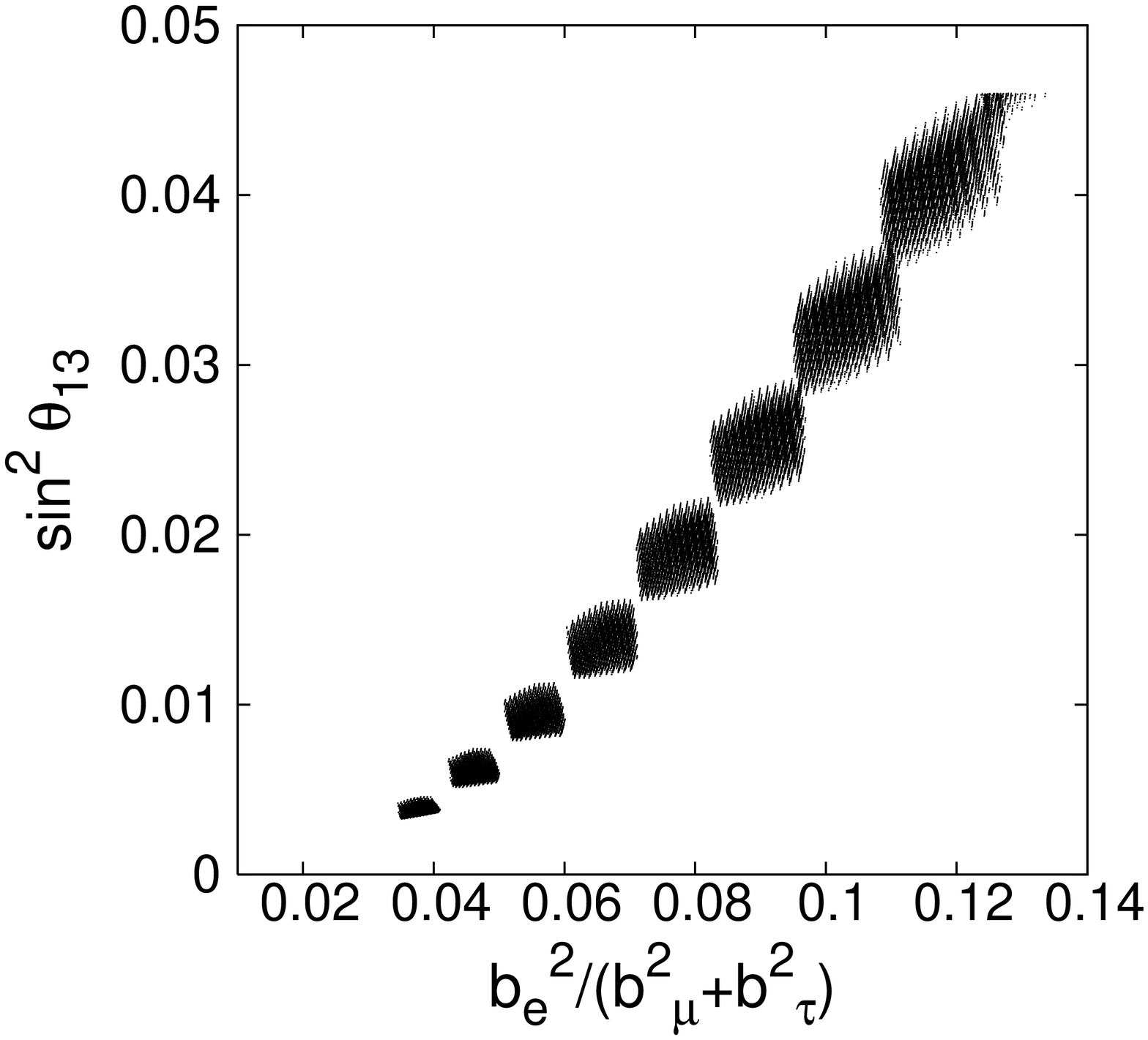
,height=8.5cm,angle=-90}
\hspace*{0mm}&\hspace*{-3mm}
\epsfig{file=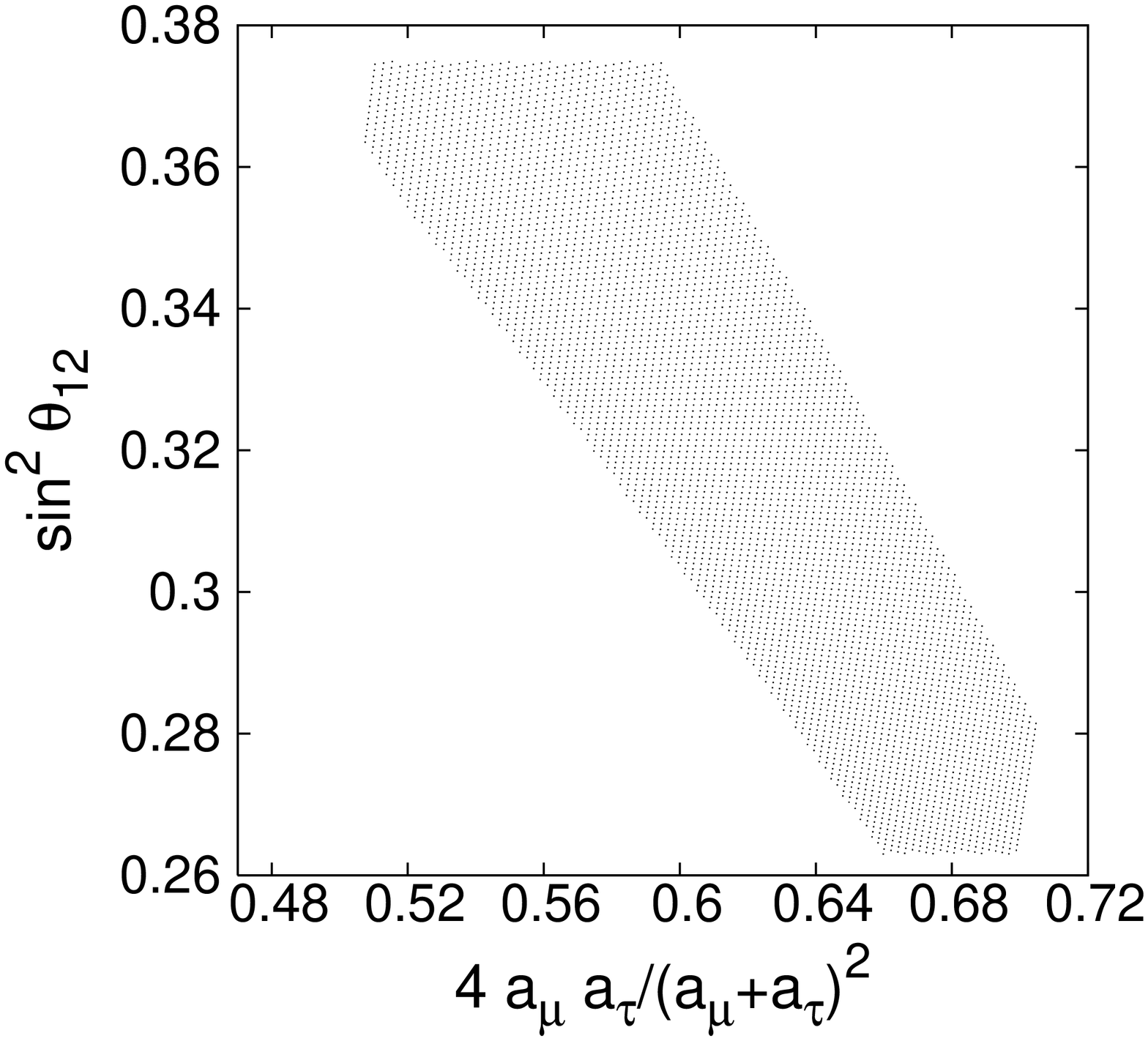,height=8.5cm,angle=-90}
      \\ & \\
      \hspace*{-1.1cm} (a) & \hspace*{-1cm} (b)
\end{tabular}
\captions{
(a) The variation of $\sin^2 \theta_{13}$ with respect to the relevant 
term that controls its evolution.
(b) The variation of $\sin^2 \theta_{12}$ with respect to the 
relevant term 
$4 a_{\mu}a_{\tau}/(a_{\mu}+a_{\tau})^2$. 
}
    \label{Seno 13 M1 positivo}
 \end{center}
\end{figure}

\begin{figure}[t]
 \begin{center}
\hspace*{-8mm}
    \begin{tabular}{cc}
    \epsfig{file=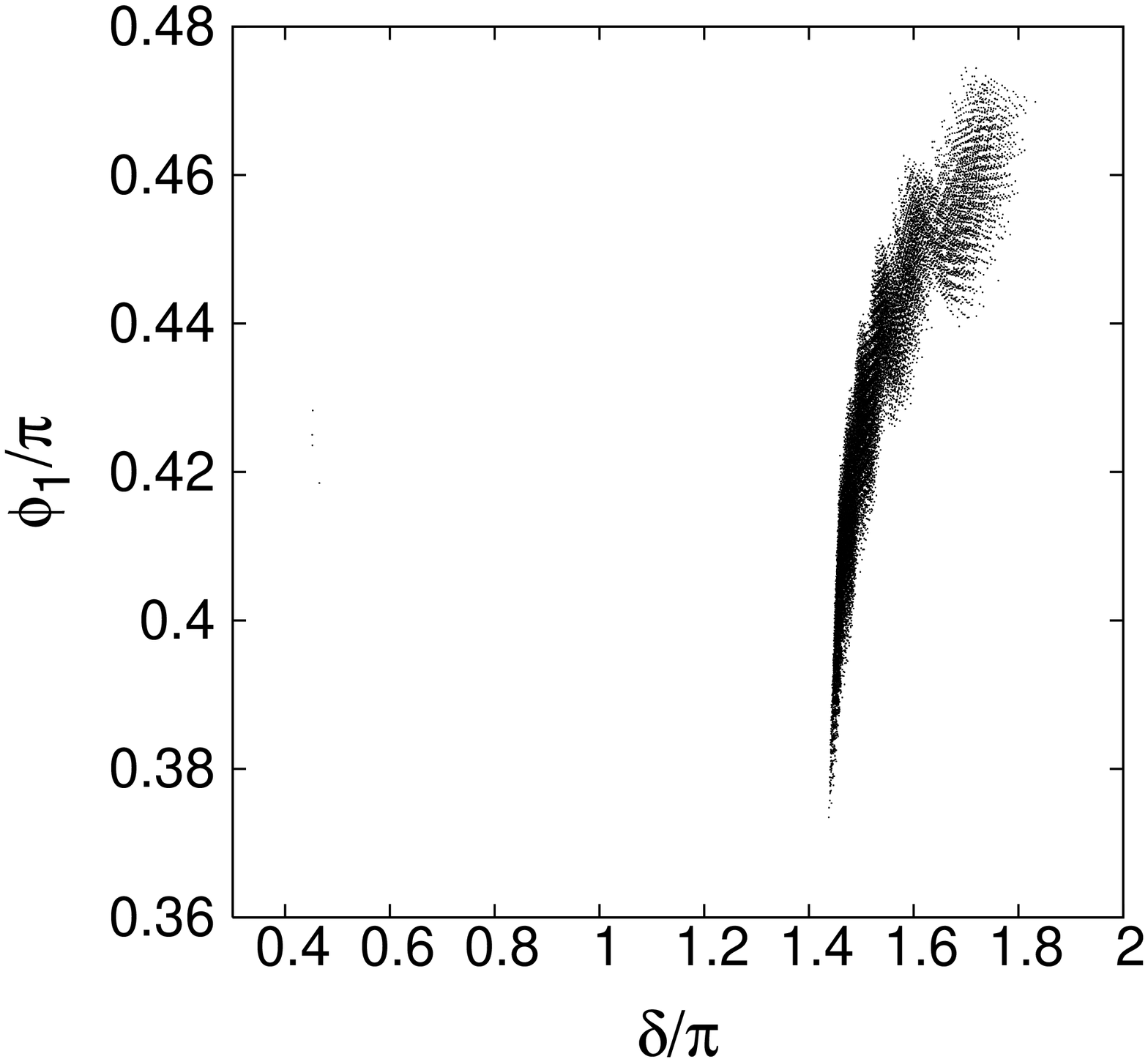
,height=8.5cm,angle=-90}
    \hspace*{0mm}&\hspace*{-3mm}
      \epsfig{file=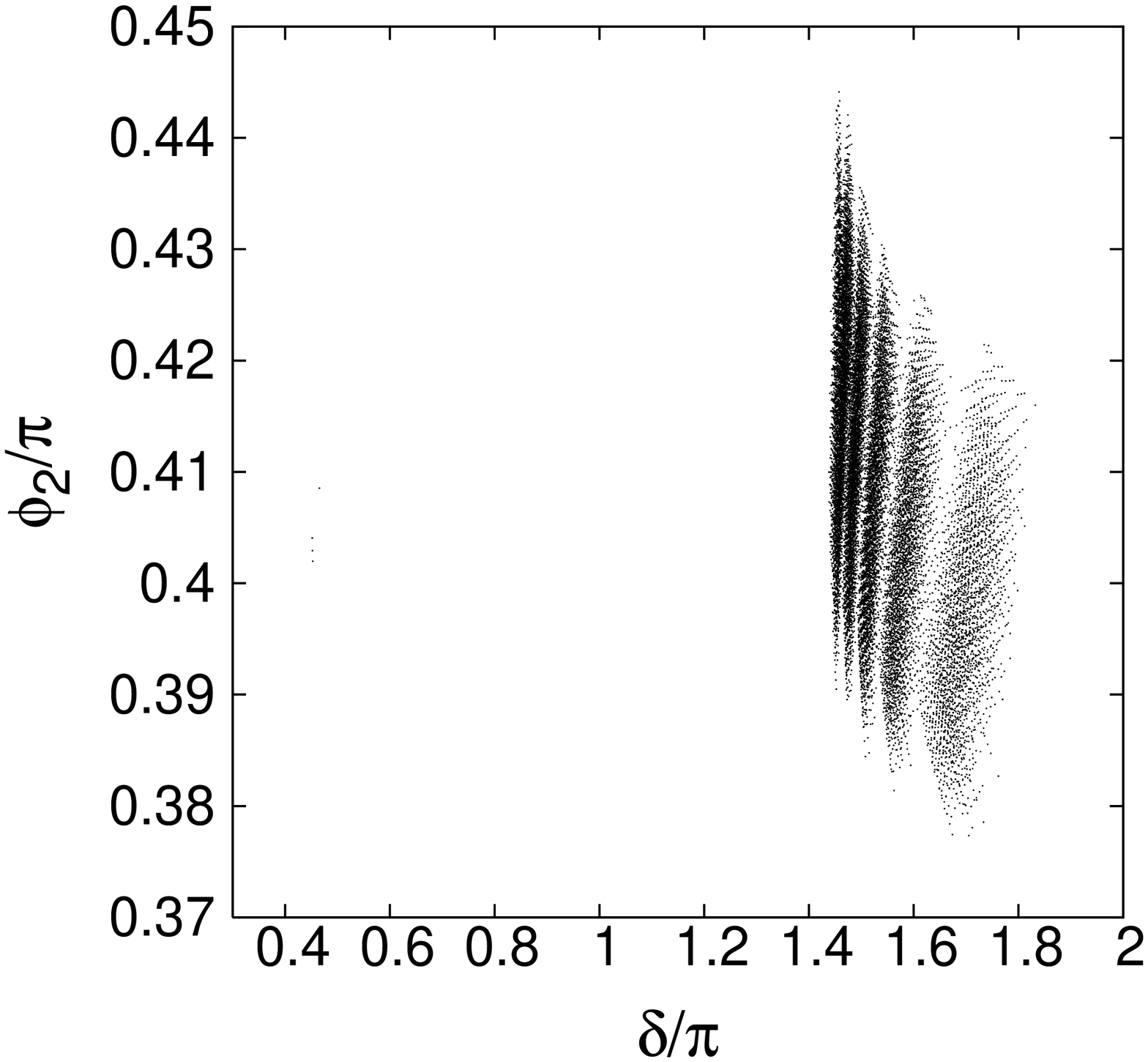,height=8.5cm,angle=-90}      \\ & \\
      \hspace*{-1.1cm} (a) & \hspace*{-1cm} (b)
    \end{tabular}
\captions{
$\delta-\phi_1$ plane (a) and $\delta-\phi_2$ plane (b)
for the scenario with normal hierarchy and positive gaugino 
masses $M>0$,
varying simultaneously 
$Y_{\nu_i}, \nu_i, M_1$. 
}
    \label{Fases CP M Pos}
 \end{center}
\end{figure}

In order to complete the discussion about the neutrino sector in this scenario, we will consider the possibility $M_1>0$ instead of $M_1<0$ .
In Sect. 3 we have seen 
that with $M_1>0$ it is more complicated to have a degeneracy between muon
and tau neutrinos because it is easy to obtain $\sin^2 \theta_{12} \sim 0$, 
in contradiction with the data (see Table \ref{Synopsis}). 
Thus we will show a region where breaking the degeneracy 
$\nu_{\mu}$-$\nu_{\tau}$ a normal hierarchy is obtained with $M_1>0$. 
This region is around the point of the parameter space shown in 
Table \ref{Punto jerarquia directa M1 postivo}. 
In this example the angle $\sin^2 \theta_{13}$ can easily be
made small as required  
by the data, but it is not necessarily negligible. Thus the CP violating effects would be present in the MNS matrix.
\begin{table}[b]
$$
\begin{array}{|c|c|c|c|}
\hline
\ Y_{\nu_1}=9.54 \times 10^{-7} \ & \ Y_{\nu_2}=9.47 \times 10^{-7} \ & \ Y_{\nu_3}=2.31 \times 10^{-7} \ & \ M_1=350 \ \textrm{GeV} \\
\hline
\ \nu_1=8.59 \times 10^{-5} \ \textrm{GeV} \ & \ \nu_2=2.25 \times 10^{-4} \ \textrm{GeV} \ & \ \nu_3=2.29 \times 10^{-4} \ \textrm{GeV} \ \\
\cline{1-3}
\end{array}
$$
\caption{
Numerical values of the relevant neutrino/neutralino-sector inputs 
that reproduce 
the neutrino experimental constraints, and correspond to the normal hierarchy 
scenario
with $M_1>0$.}
\label{Punto jerarquia directa M1 postivo}
\end{table}
Besides, 
we can roughly say that $\sin^2 \theta_{13} $ and 
$\sin^2 \theta_{12}$ are interchanged with respect to the case discussed above with $M_1 < 0$.
For completeness, in Fig. \ref{Seno 13 M1 positivo}a we show 
the variation of $\sin^2 \theta_{13}$ with respect to the term that 
controls the gaugino seesaw relevant in this case, namely  
$b_e^2/(b_{\mu}^2+b_{\tau}^2)$. 
We also plot in Fig. \ref{Seno 13 M1 positivo}b $\sin^2 \theta_{12}$ as a 
function of the relevant term that controls the Higgsino-$\nu^c$ seesaw 
$\frac{4 a_{\mu} a_{\tau}}{(a_{\mu}+ a_{\tau})^2}$.
As mention above, 
an interesting feature of this region of the parameter space is that 
the effect of the Dirac CP phase in the MNS is not removed, since the value of $\sin \theta_{13}$ is not negligible. 
Fig. \ref{Fases CP M Pos} shows the derived CP phases of the MNS matrix.

For the sake of completeness, we show in Table \ref{Punto jerarquia inversa} 
an example where the inverse hierarchy scenario is achieved.

At this point it is clear that there are many regions with different 
characteristics, different compositions for the lightest neutralino or 
regions close to the tri-bimaximal mixing regime for normal or inverted 
hierarchy that can be found with different neutrino parameters.
Furthermore, we have seen that the $\mu\nu$SSM with SCPV predicts non-zero 
CP-violating phases in the neutrino sector.
If in the future a non-zero CP violating phase in the lepton sector is measured, SCPV as the one analyzed here could be a possible source. 

Neutrino oscillations are sensitive only to the Dirac CP phase 
(insensitive to the Majorana phases). Let us briefly comment
about the possible 
determination of $\delta$ in future neutrino experiments. 
The conservation of 
CP implies 
$P(\nu_\alpha \rightarrow \nu_\beta)= 
   P(\bar \nu_\alpha \rightarrow \bar \nu_\beta)$. 
If CP is not conserved, we would have \cite{FormuladeteccionDelta}
\begin{eqnarray}
P(\nu_\mu \rightarrow \nu_e)-P(\bar \nu_\mu \rightarrow \bar \nu_e)
 =-16J \sin \left(\frac{\Delta m_{12}^2L}{4E} \right)
      \sin \left(\frac{\Delta m_{13}^2 L}{4E} \right)\sin \left(\frac{\Delta m_{23}^2 L}{4E} \right) , \nonumber \\
\label{deteccionDelta}
\end{eqnarray}
where $L$ is the oscillation length, $E$ is the neutrino beam energy 
and $J$ is the Jarlskog invariant for the neutrino mass matrix which is 
given by 
$J=s_{12}c_{12}s_{23}c_{23}s_{13}c_{13}^2\sin \delta$. There is 
only an upper experimental limit for $J$, $J<0.04$. The reason is that $J$ depends on 
$\theta_{13}$ and $\delta$, which are currently unknown. 
If $\theta_{13}$ vanishes 
(recall the bound $\sin^2 \theta_{13}<0.038$) $J$ vanishes and the effect 
of CP violation via (\ref{deteccionDelta}) would be unobservable. 
The same occurs if there was a degeneracy in the neutrino masses.
In spite of these extreme situations the process (\ref{deteccionDelta}) 
implies that long baseline experiments allow the observation of CP 
violation due to the Dirac phase $\delta$ in the neutrino sector. 
Two experiments are designed for this purpose: NO$\nu$A \cite{NOVA} and 
the T2KK detector \cite{T2KK}.

\begin{table}[t]
$$
\begin{array}{|c|c|c|c|}
\hline
\ Y_{\nu_1}=5.98 \times 10^{-7} \ & \ Y_{\nu_2}=1.32 \times 10^{-6} \ &  \ Y_{\nu_3}=1.40 \times 10^{-6} \ & \ M_1=340 \ \textrm{GeV} \\
\hline
\ \nu_1=3.276 \times 10^{-4} \ \textrm{GeV} \ & \ \nu_2=6.20 \times 10^{-5} \ \textrm{GeV} \ & \ \nu_3=6.56 \times 10^{-5} \ \textrm{GeV} \ \\
\cline{1-3}
\end{array}
$$
\caption{
Numerical values of the relevant neutrino/neutralino inputs that reproduce the neutrino experimental constraints, and correspond to the inverted hierarchy scenario.
}
\label{Punto jerarquia inversa}
\end{table}

On the other hand, although Majorana phases affect neutrinoless 
double beta decay $0\nu\beta\beta$ \cite{Neutrinoless}, their determination turn out
to be difficult. 

Let us finally briefly discuss the implications of the CP-violating phases
concerning the electric dipole moments (EDMs)
in the $\mu\nu$SSM. 
As is well known, EDMs impose important constraints on supersymmetric theories. The MSSM (with explicit CP violation in the soft Lagrangian) predicts EDMs about three orders of magnitude larger than the experimental bounds for the EDM of the electron and neutron if the supersymmetric CP violating phases are $\mathcal O(1)$ and the
supersymmetric particles have masses near their current experimental bounds $\mathcal O(100 \ GeV)$ \cite{EDMs 1}. There are three kind of solutions to this problem in supersymmetric theories. First, if the supersymmetric CP violating phases are very small, of order $\mathcal O(10^{-2}-10^{-3})$ the EDM bounds can be easily satisfied \cite{EDMs 1}. Second, if the supersymmetric scalar particles are decoupled with masses larger than about $3$ TeV, and thus out of reach of the LHC, but not spoiling the solution of supersymmetry to the hierarchy problem, the EDM bounds could also be accomplished \cite{EDMs 2}. Third, there can be internal cancellations between the different contributions to the EDMs \cite{EDMs 3}.


We would like to point out that the $\mu \nu$SSM with SCPV could implement these three kind of solutions.
First of all, the possibility of small supersymmetric CP phases is present in our model. 
Let us show for example a global minimum that break CP spontaneously with $\mathcal O(10^{-2})$ CP phases (we have also found global minima with $\mathcal O(10^{-3})$ phases).
\begin{table}[t]
$$
\begin{array}{|c|c|c|}
\hline
\ \lambda_i=0.13 \ & \ \kappa_i=0.55   \ & \ \nu_1^c=900 \ \textit{GeV} \ , \ \nu_2^c=\nu_3^c=600 \ \textit{GeV}  \ \\
\hline
\ \tan \beta=29 \ & \ \varphi_v=0 \ & \ \varphi_{\nu_1^c}=\frac{\pi}{100} \ \\
\hline
\ \varphi_{\nu_2^c}=\varphi_{\nu_3^c}=-\frac{\pi}{100} \ & \ \chi_1=-\frac{\pi}{90} \ & \ \chi_2=\chi_3=\frac{\pi}{90} \ \\
\hline
\end{array}
$$
\caption{Numerical values of the relevant inputs of a global minimum that 
breaks CP spontaneously with small phases.}
\label{Parametros del minimo global fases pequenas}
\end{table}
The values of the soft parameters not determined  by the minimization equations are chosen to be $(A_{\kappa}\kappa)_{iii}=-175$ GeV for $i \neq 1$, 
$(A_{\kappa} \kappa)_{ijk}=100$ GeV for $i,j,k \neq 1$ and 
$(A_{\kappa} \kappa)_{ijk}=-100$ GeV for one or two indices equal to $1$. The numerical values of the phases and the rest of input parameters are presented in Tables \ref{Parametros del minimo global fases pequenas} and \ref{Punto jerarquia directa M1 positiva fases pequenas}.

\begin{table}[t]
$$
\begin{array}{|c|c|c|}
\hline
\ Y_{\nu_1}=1.9 \times 10^{-7} \ & \ Y_{\nu_2}=Y_{\nu_3}=1.06 \times 10^{-6} & \ M_1=300 \ \textrm{GeV} \ \\
\hline
\ \nu_1=1.54 \times 10^{-4} \ \textrm{GeV} \ & \ \nu_2=\nu_3=2.4 \times 10^{-5} \ \textrm{GeV}  \\
\cline{1-2}
\end{array}
$$
\caption{Numerical values of the neutrino/neutralino inputs 
that reproduce the neutrino experimental constraints for the global minimum with small phases.
}
\label{Punto jerarquia directa M1 positiva fases pequenas}
\end{table}

It is worth remarking here that in models with SCPV small phases are not unnatural, since they arise as a consequence of the minimization conditions (notice 
that the use of phases as inputs in this work is just an artifact of the computation) for particular values of the soft terms. 


The other two solutions, heavy scalars and internal cancellations, can
also be implemented. Notice that the following soft parameters remain free in our model because they are not included in either the neutral scalar potential or
the neutrino sector: $(A_u Y_u)_{ij} \ , \ m^2_{\tilde u_{ij}^c} \ , \ M_3 \ , \ (A_e Y_e)_{ij} \ , \ m^2_{\tilde e^c_{ij}}$. Thus, the solution with heavy scalars remains valid for scalar masses heavier than about $3$ TeV. We also expect the internal cancellation solution to be valid in our model. This is because these free parameters enter in the calculation of the EDMs, and we will have enough freedom to find 
values where such cancellations can be accomplished, fulfilling  the EDMs bounds.



\section{Conclusions \label{section:conclusion}}
In this work we have studied in detail 
the neutrino sector of the 
$\mu\nu$SSM. 
We have also shown that, even if all parameters in the scalar potential are real, SCPV is possible at tree level, and we have used these vacua to show how a complex MNS matrix can arise.

In particular, we have calculated first the scalar potential of the $\mu\nu$SSM with real parameters, assuming the most general situation where the VEVs of Higgses and sneutrinos can be complex. We have shown, using a simple argument, that CP can actually be spontaneously violated in this model.

Then we have discussed the neutralino-neutrino mass matrix of 
the $\mu\nu$SSM, and we have shown how to obtain from it the effective 
neutrino mass matrix. Although 
the discussion is general, we have applied it also
to the
particularly interesting case of real vacua.
We have analyzed how the electroweak seesaw 
mechanism works in the $\mu\nu$SSM using approximate analytical equations, particularized 
for certain interesting limits that clarify the 
neutrino-sector behavior of the model. In addition, we have given the
qualitative idea of how to find 
regions in the parameter space of the model that satisfy the neutrino 
experimental constraints.
Let us remark that these constraints can be fulfilled  even with a diagonal neutrino Yukawa matrix, since this seesaw 
does not involve only the right-handed neutrinos but also the MSSM 
neutralinos.
Actually, to obtain the correct neutrino angles turns out to be easy
due to the following characteristics of this seesaw: R-parity is broken and the relevant scale is the electroweak one. In a sense, this gives an answer to the question why the mixing angles are so different in the quark and lepton sectors.

Finally, we have presented our results describing the method 
to obtain numerically global minima with SCPV, and 
giving examples of such minima. Let us emphasized however that, unlike the case with real VEVs where many global minima can be found, for the case with complex VEVs such minima are not so easy to find. In particular, one has to choose
carefully the parameters of the model.
For the examples found we have 
shown the dependence of the neutrino mass differences
(for both normal and inverted hierarchies), 
mixing angles, and CP phases of the MNS matrix, in terms of the relevant neutrino inputs. Last but not least,   
we have checked that different regions of the parameter space 
can reproduce the neutrino experimental constraints. 
In this context, future neutrino experiments could be able to measure a 
non-zero Dirac CP-violating phase, opening the possibility to SCPV
in the $\mu\nu$SSM as the dominant source.
\section*{Acknowledgements}\label{ack}
J. Fidalgo acknowledges the financial support of MICINN through a FPU grant. 
D.E. L\'opez-Fogliani thanks STFC for financial support. 
The work of C. Mu\~noz
was supported in part by 
MICINN under grants FPA2006-05423 and
FPA2006-01105, by the Comunidad de Madrid under grant HEPHACOS P-ESP-00346, 
and by
the European Union under the RTN program
MRTN-CT-2004-503369. The work of R. Ruiz de Austri 
was supported in part by MICINN under grant FPA2007-60323, 
by the Generalitat Valenciana under grant PROMETEO/2008/069 and by the 
Spanish Consolider-Ingenio 2010 Programme CPAN (CSD2007-00042).

\appendix
\section{Minimization Equations}
Here we write first the eight minimization conditions 
with respect to the moduli $v_d$, $v_u$, $\nu_i^c$, $\nu_i$:
\begin{align}
&
\frac{1}{4}G^2\left(\sum_i\nu_{i}\nu_{i}+v_{d}^{2}-v_{u}^{2}\right)v_{d}+m_{H_{d}}^{2}v_{d}+v_{d}v_{u}^{2}\sum_i (\lambda_{i})^2 -\sum_i (A_{\lambda} \lambda)_i \nu_i^c v_{u}\cos(\varphi_v+\varphi_{\nu_i^c})
\nonumber \\&+\sum_{i,j}v_{d} \lambda_i\lambda_{j}\nu_i^c \nu_j^c\cos(\varphi_{\nu_i^c}-\varphi_{\nu_j^c}) 
-\sum_{i,j,k}\kappa_{ikj}\lambda_{k}v_{u}\nu_i^c \nu_j^c \cos(\varphi_{\nu^c_i}+\varphi_{\nu^c_j}-\varphi_v)\nonumber \\
&-  \sum_{i,j,k}Y_{\nu_{ij}}\lambda_k \nu_{i} \nu_j^c\nu_k^c \cos(\chi_i+\varphi_{\nu_j^c}-\varphi_{\nu^c_k}-\varphi_v)-\sum_{i}\sum_{j} Y_{\nu_{ij}}\lambda_{j}v^2_{u}\nu_{i}\cos(\varphi_v-\chi_i)
=0, \label{8vd}
\end{align}
\begin{align}
&
-\frac{1}{4}G^2\left(\sum_{i}\nu_{i}\nu_{i}+v_{d}^{2}-v_{u}^{2}\right)v_{u}+m_{H_{u}}^{2}v_{u}+v_{u}v^2_{d}\sum_i (\lambda_i)^2
\nonumber \\& 
+ \sum_{i,j}(A_{\nu} Y_{\nu})_{ij}\nu_i\nu_j^c\cos(\chi_i+\varphi_{\nu^c_j})  -\sum_{i}(A_{\lambda} \lambda)_i\nu_i^c v_{d}\cos(\varphi_v +\varphi_{\nu^c_i})\nonumber \\&
+\sum_{i,j}\lambda_{i}\lambda_jv_{u}\nu_i^c\nu_j^c\cos(\varphi_{\nu^c_i}-\varphi_{\nu^c_j})-\sum_{i,j}\sum_k\kappa_{ijk}\lambda_{k}v_{d}\nu_i^c \nu_j^c \cos(\varphi_{\nu^c_i}+\varphi_{\nu_j^c}-\varphi_v)  
\nonumber \\
&+\sum_{i,j,k}\sum_lY_{\nu_{jl}}\kappa_{ilk}\nu_{j}\nu_i^c \nu_k^c \cos(\varphi_{\nu^c_i}+ \varphi_{\nu^c_k}-\chi_j) -
\sum_i \sum_j2Y_{\nu_{ij}} \lambda_j v_{d}v_{u}\nu_{i} \cos(\varphi_{v}-\chi_i)
\nonumber\\ &+\sum_{i,j}\sum_{k}Y_{\nu_{ik}}Y_{\nu_{jk}}v_{u}\nu_{i}\nu_{j}\cos(\chi_i-\chi_j)+\sum_{i,j}\sum_{k}Y_{\nu_{ki}}Y_{\nu_{kj}}v_{u}\nu_{i}^{c}\nu_j^c\cos(\varphi_{\nu_i^c}-\varphi_{\nu_j^c})=0, \label{8vu}
\end{align}
\begin{align}
&
\sum_j m^2_{\widetilde{\nu}_{ij}^{c}}\nu_{j}^{c}\cos(\varphi_{\nu^c_i}-\varphi_{\nu^c_j})-(A_{\lambda} \lambda)_i v_u v_d\cos(\varphi_v+\varphi_{\nu_i^c})+
\sum_j (A_{\nu} Y_{\nu})_{ji} \nu_{j}v_{u}\cos(\chi_j+\varphi_{\nu^c_i})
\nonumber \\ &+\sum_{j,k}(A_{\kappa} \kappa)_{ijk} \nu_{j}^{c}\nu_{k}^{c}\cos(\varphi_{\nu_i^c}+\varphi_{\nu_j^c}+\varphi_{\nu_k^c})+\sum_j\lambda_{i}\lambda_{j}v_{d}^{2}\nu_{j}^{c}\cos(\varphi_{\nu^c_i}-\varphi_{\nu^c_j}) 
\nonumber\\
&+\sum_j\lambda_i\lambda_{j}\nu_{j}^{c}v_{u}^{2}\cos(\varphi_{\nu^c_i}-\varphi_{\nu^c_j})+\sum_{j,k,l}\sum_m2\kappa_{imk}\kappa_{lmj}\nu_{j}^{c}\nu_{k}^{c}\nu_{l}^{c}\cos(\varphi_{\nu^c_i}+\varphi_{\nu^c_j}-\varphi_{\nu^c_k}-\varphi_{\nu^c_l})
\nonumber\\ &-\sum_j\sum_k2\kappa_{ijk}\lambda_{k}v_{d}v_{u}\nu_{j}^{c}\cos(\varphi_{\nu^c_i}+\varphi_{\nu^c_j}-\varphi_v)
+\sum_{j,k}\sum_{l}2Y_{\nu_{jl}}\kappa_{ikl}v_{u}\nu_{j}\nu_{k}^{c}\cos(\varphi_{\nu^c_i}+\varphi_{\nu^c_k}-\chi_j)\nonumber \\
&-\sum_{j,k}Y_{\nu_{ji}}\lambda_{k}v_{d}\nu_{j}\nu_{k}^{c}\cos(\chi_j+\varphi_{\nu^c_i}-\varphi_{\nu^c_k}-\varphi_v)-\sum_{j,k}Y_{\nu_{kj}}\lambda_{i}v_{d}\nu_{k}\nu_{j}^{c}\cos(\chi_k+\varphi_{\nu^c_j}-\varphi_{\nu^c_i}-\varphi_v)\ 
\nonumber \\
&+\sum_{j,k,l}Y_{\nu_{ji}}Y_{\nu_{lk}}\nu_{j}\nu_{l}\nu_{k}^{c}\cos(\chi_j-\chi_k+\varphi_{\nu^c_i}-\varphi_{\nu^c_l})+\sum_{j}\sum_{k}Y_{\nu_{ki}}Y_{\nu_{kj}}v_{u}^{2}\nu^c_{j}\cos(\varphi_{\nu^c_i}-\varphi_{\nu^c_j}) =0, \label{8r} \end{align}
\begin{align}
&
\frac{1}{4}G^2(\sum_j \nu_{j}\nu_{j}+v_{d}^{2}-v_{u}^{2})\nu_{i}
+\sum_j m^2_{\tilde{L}_{ij}}\nu_{j}\cos(\chi_i-\chi_j)+\sum_{j} (A_{\nu} Y_{\nu})_{ij} \nu_{j}^{c}v_{u}\cos(\chi_i+\varphi_{\nu_j^c}) \nonumber \\
&+\sum_{j,k}\sum_{l}Y_{\nu_{il}}\kappa_{jlk}v_{u}\nu^c_{j}\nu^c_{k}\cos(\varphi_{\nu^c_j}+\varphi_{\nu_k^c}-\chi_i) \nonumber \\&-\sum_{j,k}Y_{\nu_{ij}} \lambda_{k}v_{d}\nu_j^c\nu_k^c \cos(\chi_i+\varphi_{\nu_j^c}-\varphi_{\nu^c_k}-\varphi_v)-\sum_{j}Y_{\nu_{ij}}\lambda_{j}v_{d}v_{u}^{2} \cos(\varphi_{v}-\chi_i)\nonumber\\
&+\sum_{j,k,l}Y_{ij}
Y_{\nu_{kl}}\nu_j^c\nu_{k}\nu_l^c \cos(\chi_i-\chi_k+\varphi_{\nu^c_j}-\varphi_{\nu^c_l})
+\sum_{j}\sum_{k}Y_{\nu_{ik}}Y_{\nu_{jk}}v_{u}^{2} \nu_j \cos({\chi_i-\chi_j})\nonumber \\&
=0. \label{8n}
\end{align}
The seven minimization conditions 
with respect to the phases $\varphi_v$, $\varphi_{\nu^c_i}$ and
$\chi_i$ are:
\begin{eqnarray}
&-&\sum_{i,j}\sum_k2\kappa_{ijk}\lambda_{k}v_{d}v_{u}\nu^c_i\nu_{j}^{c}\sin(\varphi_{\nu^c_i}+\varphi_{\nu^c_j}-\varphi_v)\nonumber\\
&-&2[\sum_{i,j,k}Y_{\nu_{ij}}\lambda_{k}v_{d}{\nu}_{i}{\nu}^c_{j}{\nu}^c_{k} \sin(\chi_i+\varphi_{\nu^c_j}-\varphi_{\nu^c_k}-\varphi_{v})
-\sum_{i}\sum_{j}Y_{\nu_{ij}}\lambda_{j}v_{d}v^2_{u}{\nu}_{i}\sin(\varphi_{v}-\chi_i)]\nonumber\\
&+&2\sum_{i} (A_{\lambda} \lambda)_i {\nu}^c_{i}v_{d}v_{u} \sin(\varphi_{v} +\varphi_{\nu^c_i})=0,\label{7v}
\end{eqnarray}
\begin{eqnarray}
&-&\sum_{j} m_{\tilde{\nu}^c_{ij}}^{2}{\nu}^c_{i}{\nu}^c_{j} \sin(\varphi_{\nu^c_i}-\varphi_{\nu^c_j})\nonumber\\
 &-&\sum_{j}\lambda_{i}\lambda_{j}v_{d}^2{\nu}^c_{i}{\nu}^{c}_{j}\sin(\varphi_{\nu^c_i}-\varphi_{\nu^c_j})
-\sum_{j}\lambda_{i}\lambda_{j}v_{u}^2{\nu}^c_{i}{\nu}^{c}_{j}\sin(\varphi_{\nu^c_i}-\varphi_{\nu^c_j})\nonumber\\
&-&2\sum_{j,k,l}\sum_{m}
\kappa_{imk}\kappa_{lmj}{\nu}^c_{i}{\nu}^{c}_{j}{\nu}^c_{k}{\nu}^{c}_{l}\sin(\varphi_{\nu^c_i}+\varphi_{\nu^c_j}-\varphi_{\nu^c_k}-\varphi_{\nu^c_l})
\nonumber\\
&+&2\sum_{j,k}
\kappa_{ikj}\lambda_{k}v_{d}v_{u}{\nu}^c_{i}{\nu}^c_{j}\sin(\varphi_{\nu^c_i}+\varphi_{\nu^c_j}-\varphi_v)\ \nonumber\\
&-&2\sum_{j,k}\sum_{l}
Y_{\nu_{jl}}\kappa_{ilk}v_{u}{\nu}_{j}{\nu}^c_{i}{\nu}^c_{k}\sin(\varphi_{\nu^c_i}+\varphi_{\nu^c_k}-\chi_j)\ 
\nonumber\\&+&\sum_{j,k}Y_{\nu_{ji}}\lambda_{k}v_{d}{\nu}_{j}{\nu}^c_{i}{\nu}^c_{k} \sin(\chi_j+\varphi_{\nu^c_i}-\varphi_{\nu^c_k}-\varphi_{v})-\sum_{j,k}Y_{\nu_{kj}}\lambda_{i}v_{d}{\nu}_{k}{\nu}^c_{j}{\nu}^c_{i} \sin(\chi_k+\varphi_{\nu^c_j}-\varphi_{\nu^c_i}-\varphi_{v})
\nonumber \\&-&\sum_{j,k,l}Y_{\nu_{ji}}Y_{\nu_{kl}}{\nu}_{j}{\nu}^c_{i}{\nu}_{k}{\nu^{c}_{l}} \sin(\chi_j-\chi_k+\varphi_{\nu^c_i}-\varphi_{\nu^c_l})-\sum_{j}\sum_{k}Y_{\nu_{ki}} Y_{\nu_{kj}}v^2_{u} {\nu}^c_{i}{\nu^{c}_{j}}\sin(\varphi_{\nu_i^c}-\varphi_{\nu_j^c}) \
\nonumber\\
&+& (A_{\lambda} \lambda)_i {\nu}^c_{i}v_{d}v_{u} \sin(\varphi_{v} +\varphi_{\nu^c_i})-\sum_{j,k}  ({A_{\kappa} \kappa)_{ijk}{\nu}^c_{i}{\nu}^c_{j}{\nu}^c_{k} \sin(\varphi_{\nu^c_i}+\varphi_{\nu^c_j}+\varphi_{\nu^c_k})}
\nonumber \\ &-& \sum_{j} (A_{\nu} Y_{\nu})_{ji} v_{u}{\nu}_{j}{\nu}^{c}_{i} \sin(\chi_j+\varphi_{\nu^c_i})=0,
\label{7r}
\end{eqnarray}
\begin{eqnarray}
&-&\sum_{j} m_{\tilde{L}_{ij} }^2 \, {\nu_i} \, {\nu_j} \sin(\chi_i-\chi_j) \nonumber\\
&+&\sum_{j,k}\sum_{l}
Y_{\nu_{il}}\kappa_{jlk}v_{u}{\nu}_{i}{\nu}^c_{j}{\nu}^c_{k}\sin(\varphi_{\nu^c_j}+\varphi_{\nu^c_k}-\chi_i)+\sum_{j,k}Y_{\nu_{ij}}\lambda_{k}v_{d}{\nu}_{i}{\nu}^c_{j}{\nu}^c_{k} \sin(\chi_i+\varphi_{\nu^c_j}-\varphi_{\nu^c_k}-\varphi_{v})\ 
\nonumber\\
&-&
\sum_{j}Y_{\nu_{ij}}\lambda_{j}v_{d}v^2_{u}{\nu}_{i}\sin(\varphi_{v}-\chi_i)\nonumber\\
&-&\sum_{j,k,l}Y_{\nu_{ij}}Y_{\nu_{kl}}{\nu}_{i}{\nu}^c_{j}{\nu}_{k}{\nu^{c}_{l}} \sin(\chi_i-\chi_k+\varphi_{\nu^c_j}-\varphi_{\nu^c_l})\nonumber \\&-&\sum_{j}\sum_{k}Y_{\nu_{ik}}Y_{\nu_{jk}}v^2_{u}{\nu}_{i}{\nu}_{j}\sin(\chi_i-\chi_j) \
\nonumber\\
&-& \sum_{j} (A_{\nu} Y_{\nu})_{ij} v_{u}{\nu}_{i}{\nu}^{c}_{j} \sin(\chi_i+\varphi_{\nu^c_j})=0.
\label{7n}
\end{eqnarray}
\section{Analitical formula for neutrino masses} \label{formulote}
The formula presented here is obtained from Eq. (\ref{eff}) neglecting terms 
proportional to $Y_{\nu}^2 \nu^2$, $Y_{\nu}^3 \nu$
and $Y_{\nu} \nu^3$, and has been 
particularized for the simplified case discussed in Sect. 4 of a common value of couplings 
$\lambda_i \equiv \lambda$, 
a tensor $\kappa$ with terms $\kappa_{iii} \equiv \kappa_i \equiv \kappa$ and vanishing otherwise, diagonal Yukawa couplings $Y_{\nu_{ii}}\equiv  Y_{\nu_i}$,
and a common value of the VEVs $\nu_i^c\equiv\nu^c$. 
The phase structure of the
global minimum discussed in Section 4 for analyzing the neutrino sector,
$\varphi_{\nu_1^c}=-\varphi_{\nu_2^c}=-\varphi_{\nu_3^c} \equiv \varphi_{\nu^c}$ 
and 
$\varphi_{\nu_1}=-\varphi_{\nu_2}=-\varphi_{\nu_3} \equiv \varphi_{\nu}$,
has also been used in the computation.
Then we arrive to the following formula:
\begin{eqnarray}
(m_{eff})_{ij} \simeq \frac{X_{ij}}{\Delta}+\frac{T_{ij}}{Z} \ \frac{a_i a_j}{2 \kappa \nu^c},
\label{Analytical approximate effective neutrino mass matrix}
\end{eqnarray}
where the parameters have been defined as
\begin{eqnarray}
a_i &=& Y_{\nu_i} v_u, \nonumber \\
\Delta &=& (e^{i \varphi_{\nu^c}}+2 e^{i 3 \varphi_{\nu^c}}) \lambda^2 (v_u^2+v_d^2)^2+(8 e^{i \varphi_{\nu^c}}+4 e^{i 3 \varphi_{\nu^c}}) \lambda \kappa {\nu^c}^2 v_d v_u e^{-i \varphi_v} \nonumber \\ 
& -& (16+16 e^{i 2 \varphi_{\nu^c}}+4 e^{i 4 \varphi_{\nu^c}})M \lambda^2 \kappa {\nu^c}^3-(8+20 e^{i 2 \varphi_{\nu^c}}+8 e^{i 4 \varphi_{\nu^c}})M \lambda^3 \nu^c v_d v_u e^{i \varphi_v},  \nonumber \\
Z &=& e^{i \varphi_{\nu^c}} [-4 e^{i \varphi_{\nu^c}}(2+e^{i 2 \varphi_{\nu^c}})\kappa {\nu^c}^2 v_d v_u+e^{i \varphi_v}\lambda (4 M(2+e^{i 2 \varphi_{\nu^c}})^2 \kappa {\nu^c}^3 \nonumber \\
& -& e^{i \varphi_{\nu^c}}(1+2 e^{i 2 \varphi_{\nu^c}})(v_d^2+v_u^2)^2)+4 e^{i 2 (\varphi_{\nu^c}+\varphi_v)}\lambda^2 M \nu^c v_d v_u(5+4 \cos2 \varphi_{\nu^c}) ],
\label{Parameters of the approximate analytical expression}
\end{eqnarray}
with $\frac{1}{M} = \frac{g_1^2}{M_1}+\frac{g_2^2}{M_2}$,
\begin{eqnarray}
T_{11} &=& 2 e^{i 2 \varphi_v}[-4e^{i2(\varphi_{\nu^c}+\varphi_v)}(2+e^{i2\varphi_{\nu^c}})M \lambda^2 \nu^c v_d v_u+4e^{i\varphi_{\nu^c}}\kappa \nu^{c^2}v_d v_u \nonumber \\
&+& e^{i \varphi_v}\lambda(-4(2+e^{i6 \varphi_{\nu^c}})M\kappa\nu^{c^3}+e^{i3\varphi_{\nu^c}}(v_u^2+v_d^2)^2)], \nonumber \\
T_{22} &=& T_{33}=2 e^{i (\varphi_{\nu^c}+2 \varphi_v)}[-4e^{i2(\varphi_{\nu^c}+\varphi_v)}(2+e^{i2\varphi_{\nu^c}})M \lambda^2 \nu^c v_d v_u+4e^{i3 \varphi_{\nu^c}}\kappa \nu^{c^2}v_d v_u \nonumber \\
&+& e^{i \varphi_v}\lambda(-4(1+e^{i2 \varphi_{\nu^c}}+e^{i4\varphi_{\nu^c}})M\kappa\nu^{c^3}+e^{i3\varphi_{\nu^c}}(v_u^2+v_d^2)^2)], \nonumber \\
T_{12} &=& T_{13}=- e^{i 2 \varphi_v}[-4e^{i2(\varphi_{\nu^c}+\varphi_v)}(2+e^{i2\varphi_{\nu^c}})M \lambda^2 \nu^c v_d v_u+4e^{i3\varphi_{\nu^c}}\kappa \nu^{c^2}v_d v_u \cos(2 \varphi_{\nu^c}) \nonumber \\
&+& e^{i (3\varphi_{\nu^c}+\varphi_v)}\lambda(4(-3 \cos(3\varphi_{\nu^c})+i \sin{(3\varphi_{\nu^c}}))M\kappa\nu^{c^3}+(v_u^2+v_d^2)^2)], \nonumber \\
T_{23} &=& - e^{i 2 \varphi_v}[-4e^{i2(2\varphi_{\nu^c}+\varphi_v)}(2+e^{i2\varphi_{\nu^c}})M \lambda^2 \nu^c v_d v_u+4e^{i3\varphi_{\nu^c}}\kappa \nu^{c^2}v_d v_u  \nonumber \\
&+& e^{i \varphi_v}\lambda(-4(-1+4e^{i3\varphi_{\nu^c}}\cos(\varphi_{\nu^c}))M\kappa\nu^{c^3}+e^{i5\varphi_{\nu^c}}(v_u^2+v_d^2)^2)],
\label{Parameters of the approximate analytical expression 2}
\end{eqnarray}
and
\begin{eqnarray}
X_{11} &=& 2 \kappa \nu^{c^3}(b_{11})^2+2 \lambda \nu^c v_d v_u e^{i \varphi_v}(b'_{11})^2+\epsilon_{11}, \nonumber \\
X_{22} &=& 2 \kappa \nu^{c^3}(b_{22})^2+2 \lambda \nu^c v_d v_u e^{i \varphi_v}(b'_{22})^2+\epsilon_{22}, \nonumber \\
X_{33} &=& 2 \kappa \nu^{c^3}(b_{33})^2+2 \lambda \nu^c v_d v_u e^{i \varphi_v}(b'_{33})^2+\epsilon_{33}, \nonumber \\
X_{12} &=& 2 \kappa \nu^{c^3}(b_{11})(b_{22})+2 \lambda \nu^c v_d v_u e^{i \varphi_v}(b'_{12})^2+\epsilon_{12}, \nonumber \\
X_{13} &=& 2 \kappa \nu^{c^3}(b_{11})(b_{33})+2 \lambda \nu^c v_d v_u e^{i \varphi_v}(b'_{13})^2+\epsilon_{13}, \nonumber \\
X_{23} &=& 2 \kappa \nu^{c^3}(b_{22})(b_{33})+2 \lambda \nu^c v_d v_u e^{i \varphi_v}(b'_{23})^2+\epsilon_{23}, 
\label{Parameters of the approximate analytical expression 4}
\end{eqnarray}
with
\begin{eqnarray}
(b_{11}) &=& (2+e^{i2 \varphi_{\nu^c}})\lambda e^{-i \varphi_\nu}\nu_1+e^{i2\varphi_{\nu^c}}v_d Y_{\nu_1}, \nonumber \\
(b_{22}) &=& (2+e^{i2 \varphi_{\nu^c}})\lambda e^{i \varphi_\nu}\nu_2+v_d Y_{\nu_2}, \nonumber \\
(b_{33}) &=& (2+e^{i2 \varphi_{\nu^c}})\lambda e^{i \varphi_\nu}\nu_3+v_d Y_{\nu_3}, \nonumber \\
(b'_{11})^2 &=& (2+5e^{i2\varphi_{\nu^c}}+2e^{i4\varphi_{\nu^c}})\lambda^2e^{-i2\varphi_\nu}\nu_1^2 \nonumber \\
&+& (2+2e^{i2\varphi_{\nu^c}}+2e^{i4\varphi_{\nu^c}})\lambda v_d e^{-i \varphi_\nu}\nu_1 Y_{\nu_1}+e^{i2\varphi_{\nu^c}}v_d^2 Y_{\nu_1}^2, \nonumber \\
(b'_{22})^2 &=& (2+5e^{i2\varphi_{\nu^c}}+2e^{i4\varphi_{\nu^c}})\lambda^2 e^{i2\varphi_\nu}\nu_2^2 \nonumber \\
&+& (1+4e^{i2\varphi_{\nu^c}}+e^{i4\varphi_{\nu^c}})\lambda v_d e^{i\varphi_\nu}\nu_2 Y_{\nu_2}+e^{i2\varphi_{\nu^c}}v_d^2 Y_{\nu_2}^2, \nonumber \\
(b'_{33})^2 &=& (2+5e^{i2\varphi_{\nu^c}}+2e^{i4\varphi_{\nu^c}})\lambda^2 e^{i2\varphi_\nu}\nu_3^2 \nonumber \\
&+& (1+4e^{i2\varphi_{\nu^c}}+e^{i4\varphi_{\nu^c}})\lambda v_d e^{i\varphi_\nu}\nu_3 Y_{\nu_3}+e^{i2\varphi_{\nu^c}}v_d^2 Y_{\nu_3}^2, \nonumber \\
(b'_{12})^2 &=& (2+5e^{i2\varphi_{\nu^c}}+2e^{i4\varphi_{\nu^c}})\lambda^2 \nu_1 \nu_2+(1+e^{i2\varphi_{\nu^c}}+e^{i4\varphi_{\nu^c}})\lambda v_d e^{i\varphi_{\nu}}\nu_2 Y_{\nu_1} \nonumber \\
&+& ((1/2)+2e^{i2\varphi_{\nu^c}}+(1/2)e^{i4\varphi_{\nu^c}})\lambda v_d e^{-i \varphi_{\nu}}\nu_1 Y_{\nu_2}+(1/2)(1+e^{i4\varphi_{\nu^c}})v_d^2 Y_{\nu_1}Y_{\nu_2}, \nonumber \\
(b'_{13})^2 &=& (2+5e^{i2\varphi_{\nu^c}}+2e^{i4\varphi_{\nu^c}})\lambda^2 \nu_1 \nu_3+(1+e^{i2\varphi_{\nu^c}}+e^{i4\varphi_{\nu^c}})\lambda v_d e^{i\varphi_{\nu}}\nu_3 Y_{\nu_1} \nonumber \\
&+& ((1/2)+2e^{i2\varphi_{\nu^c}}+(1/2)e^{i4\varphi_{\nu^c}})\lambda v_d e^{-i \varphi_{\nu}}\nu_1 Y_{\nu_3}+(1/2)(1+e^{i4\varphi_{\nu^c}})v_d^2 Y_{\nu_1}Y_{\nu_3}, \nonumber \\
(b'_{23})^2 &=& (2+5e^{i2\varphi_{\nu^c}}+2e^{i4\varphi_{\nu^c}})\lambda^2 e^{i2\varphi_{\nu}}\nu_2 \nu_3 \nonumber \\
&+& ((1/2)+2e^{i2\varphi_{\nu^c}}+(1/2)e^{i4\varphi_{\nu^c}})\lambda v_d e^{i\varphi_\nu}(\nu_3 Y_{\nu_2}+\nu_2 Y_{\nu_3})+e^{i2\varphi_{\nu^c}}v_d^2 Y_{\nu_2}Y_{\nu_3}
\label{Parameters of the approximate analytical expression 5}
\end{eqnarray}
and
\begin{eqnarray}
\epsilon_{11} &=& (4 e^{i4 \varphi_{\nu^c}}-4)\lambda^2 \nu^c v_u^3 e^{i\varphi_v}e^{-i\varphi_\nu}\nu_1 Y_{\nu_1}, \nonumber \\
\epsilon_{22} &=& (2-2e^{i4\varphi_{\nu^c}})\lambda^2 \nu^c v_u^3 e^{i\varphi_v}e^{i\varphi_\nu}\nu_2 Y_{\nu_2}, \nonumber \\
\epsilon_{33} &=& (2-2e^{i4\varphi_{\nu^c}})\lambda^2 \nu^c v_u^3 e^{i\varphi_v}e^{i\varphi_\nu}\nu_3 Y_{\nu_3}, \nonumber \\
\epsilon_{12} &=& (2e^{i4\varphi_{\nu^c}}-2)\lambda^2 \nu^c v_u^3 e^{i\varphi_v}e^{i\varphi_\nu}\nu_2 Y_{\nu_1}+(1-e^{i4\varphi_{\nu^c}})\lambda^2 \nu^c v_u^3 e^{i\varphi_v}e^{-i\varphi_{\nu}}\nu_1 Y_{\nu_2}, \nonumber \\
\epsilon_{13} &=& (2e^{i4\varphi_{\nu^c}}-2)\lambda^2 \nu^c v_u^3 e^{i\varphi_v}e^{i\varphi_\nu}\nu_3 Y_{\nu_1}+(1-e^{i4\varphi_{\nu^c}})\lambda^2 \nu^c v_u^3 e^{i\varphi_v}e^{-i\varphi_{\nu}}\nu_1 Y_{\nu_3}, \nonumber \\
\epsilon_{23} &=& (1-e^{i4\varphi_{\nu^c}})\lambda^2 \nu^c v_u^3 e^{i\varphi_v}e^{i\varphi_\nu}(\nu_3 Y_{\nu_2}+\nu_2 Y_{\nu_3}). 
\label{Parameters of the approximate analytical expression 6}
\end{eqnarray}
Let us discuss two particular limits where the formula becomes simple.
In the limit $M \to \infty$ and $v_d \to 0$ we obtain
\begin{eqnarray}
(m_{eff})_{ij} \simeq F_{ij} \ \frac{a_i a_j}{2 \kappa \nu^c},
\label{Analytical approximate effective neutrino mass matrix limit}
\end{eqnarray}
where 
\begin{eqnarray}
F_{11} &=& -2 e^{i (2 \varphi_v-\varphi_{\nu^c})} (2+e^{i6 \varphi_{\nu^c}})\left(2+e^{2i \varphi_{\nu^c}} \right)^{-2}, \nonumber \\
F_{22} &=& F_{33}=-2 e^{i (2 \varphi_v-\varphi_{\nu^c})} (1+e^{i2 \varphi_{\nu^c}}+e^{i4\varphi_{\nu^c}}) \left(2+e^{2i \varphi_{\nu^c}} \right)^{-2}, \nonumber \\
F_{12} &=& F_{13}=e^{i 2(\varphi_{\nu^c}+\varphi_v)}(3 \cos(3\varphi_{\nu^c})-i \sin{(3\varphi_{\nu^c}}) \left(2+e^{2i \varphi_{\nu^c}} \right)^{-2} , \nonumber \\
F_{23} &=& e^{i (2 \varphi_v-\varphi_{\nu^c})}(4e^{i3\varphi_{\nu^c}}\cos(\varphi_{\nu^c})-1) \left(2+e^{2i \varphi_{\nu^c}}\right)^{-2}.
\label{Parameters of the approximate analytical expression 2 facilita}
\end{eqnarray}
In the limit of vanishing phases, i.e. 
real VEVs, we obtain
\begin{eqnarray}
(m_{eff|real})_{ij} & \simeq & \frac{2}{3}\frac{(\kappa \nu^{c^2}+\lambda v_u v_d)\nu^c }{\lambda^2 (v_u^2+v_d^2)^2+4 \lambda \kappa \nu^{c^2}v_u v_d-12M\lambda (\kappa \nu^{c^2}+\lambda v_u v_d)\lambda \nu^c}\ b_i b_j \nonumber \\
& + & \frac{1}{6 \kappa \nu^c}(1-3 \delta_{ij})a_i a_j,
\label{Formula analitica real}
\end{eqnarray}
where we have defined
\bea
 b_i=Y_{\nu_i}v_d+3 \lambda \nu_i.
\eea
Regarding the previous parameters we note that for the real case
\bea
b_i=b_{ii}=b'_{ii}, \nonumber \\
b'^2_{ij}=b_{ii}\,b_{jj}=b_i \, b_j, \nonumber \\
\epsilon_{ij}=0.
\eea

\end{document}